\documentclass{aa}  

\usepackage{xcolor}
\usepackage{graphicx}
\usepackage{txfonts}
\usepackage[normalem]{ulem}
\usepackage{lipsum}
\usepackage{subcaption}         
\usepackage{lscape}             
\usepackage{placeins}           
 \usepackage{comment}                               
\newcommand{\ep}{{\it EP}}
\newcommand{\leia}{{\it LEIA}}

\begin{document}

   \title{X-ray Activity of the RS CVn-type Star $\sigma$ Gem with the First-Year Observations of Einstein Probe}


   \author{X. Mao\inst{1,2,3}\and G. Micela\inst{3}\and F. Favata\inst{3,4} 
    \and W. Yuan\inst{1,2} \and H. Liu\inst{1} \and H. Cheng\inst{1}
        }

    \institute{National Astronomical Observatories, Chinese Academy of Sciences, 100101, Beijing, China\\\email{maoxuan@bao.ac.cn}\\
    \and School of Astronomy and Space Science, University of Chinese Academy of Sciences, Chinese Academy of Sciences, 100049, Beijing, China\\
    \and INAF - Osservatorio Astronomico di Palermo, Piazza del Parlamento 1, 90134 Palermo, Italy\\\email{giuseppina.micela@inaf.it}
    \and Department of Physics, Imperial College London, Exhibition Road, London SW7 2AZ, UK}

   \date{Received September 30, 20xx}

 
  \abstract
   {Stellar flares are energetic events driven by the sudden release of magnetic energy in the stellar atmosphere. Compared to the Sun, the diversity of stellar types and environments leads to a richer variety of flaring phenomena. Studying these flares is crucial for understanding their impact on exoplanets, the circumstellar environment, and stellar evolution itself. The launch of the Einstein Probe (\ep) offers a unique opportunity to systematically detect such events.}
   {We present a systematic analysis of the flaring activity of the active RS CVn-type binary $\sigma$ Gem, utilizing the first-year monitoring data from the Wide-field X-ray Telescope (WXT) aboard \ep. Our goals are to demonstrate the unique capability of \ep\ in monitoring stellar X-ray activity and detecting flares, by identifying and characterizing extreme X-ray flares on $\sigma$ Gem and estimating their occurrence rate.} 
   {We developed a data‑processing pipeline to select and extract \ep-WXT observations, producing a background-subtracted, vignetting-corrected light curve. We employed the Bayesian Blocks method to detect significant flares in the long-term X-ray light curve. The key detection parameter was optimized through extensive simulations. For each identified flare, we performed light curve and spectral fitting to derive its duration, peak luminosity, total energy, and plasma properties (e.g., temperature, emission measure).}
   {Between October 2024 and April 2025, WXT detected 6 distinct flares from $\sigma$ Gem. Their durations ranged from 21 hours to 3 days, with peak X-ray luminosities (0.5--4 keV) of $3.7\times10^{31}$ to $7.0\times10^{32}$ erg s$^{-1}$ and total energies of $1.1\times10^{36}$ to $4.4\times10^{37}$ erg, placing them among the “superflare” class.}
   {Using $\sigma$ Gem as a case study, we demonstrate an analysis process for flare detection and analysis with \ep-WXT data, which has rapidly expanded the catalog of known X-ray flares for this star and provides new statistical constraints on its flaring behavior. Applying this methodology to the growing \ep\ stellar archive promises to yield a vast sample of X-ray flares, which will significantly advance our understanding of stellar magnetic activity.}


   \maketitle
   \nolinenumbers

\section{Introduction}
Stellar flares are transient energetic phenomena characterized by the sudden release of a large amount of energy from localized regions in the extended stellar atmosphere, producing emission across the electromagnetic spectrum from radio to gamma-ray. By analogy with solar flares, stellar flares are commonly interpreted as being powered by magnetic reconnection processes in stellar coronae \citep{Yokoyama1995, Reale1998}. Observationally, flaring activity has been detected in a wide range of stellar types \citep{Favata2000, Pandey2012, Schmidt2016, Getman2021F, Vasilyev2024}. Compared to the Sun, stars span a broader range of physical properties and environments, resulting in more complex magnetic field geometries and a greater diversity of flare characteristics \citep{Favata2000,Pandey2008,Schmidt2016,Getman2021F,Karmakar2023}.

\begin{table*}[h!]
\caption{Specifications of the Instruments.}
\label{tab:instru}
\centering
\begin{tabular}{ccc}
\hline\hline
Parameters & WXT & FXT \\
\hline
Telescope optic & lobster-eye MPO & Wolter-I \\
Detector & CMOS & pn-CCD \\
Field of view & 3850 deg$^{2}$ & $\geq$60' (diameter) \\
Effective area @1.25 keV (cm$^2$) & 2--3 & $\sim$300 (one unit) \\
Spatial resolution (1 keV) & 5' (FWHM) & 20--24" (HPD, on-axis) \\
Bandpass (keV) & 0.5--4 & 0.5--10 \\
Limiting flux (erg s$^{-1}$ cm$^{-2}$) & $\sim$2.6$\times$10$^{-11}$ @1 ks & $\sim$10$^{-14}$ @10 ks \\
Energy resolution (eV) & 122 @1.25 keV & 100 @1.5 keV \\
 &  & 50 ms (full-frame) \\
Time resolution & 50 ms & 2 ms (partial window)\\
 &  & 42 $\mu$s (timing) \\
\hline
\end{tabular}
\tablefoot{MPO: micro-pore optics; FWHM: full width at half maximum; HPD: half-power diameter.}
\end{table*}

However, systematic studies of stellar X‑ray flares, particularly extreme, high‑luminosity events, have long been hampered by observational challenges. Historically, the detection of such events has been sparse, largely due to the scarcity of X‑ray instruments that combine a wide field of view with high sensitivity. Studies have thus relied on two complementary strategies, each with inherent limitations. On the one hand, all‑sky monitors such as {\it MAXI}/GSC \citep{Matsuoka2009} and {\it Swift}/BAT \citep{Barthelmy2005} can capture only the most luminous flares \citep{Osten2010, Tsuboi2016, Sasaki2021, Karmakar2023}, thus missing the bulk of the flare population. 
On the other hand, pointed observatories such as {\it XMM‑Newton} \citep{Jansen2001} and {\it Chandra} \citep{Weisskopf2002} offer high sensitivity and have detected numerous flares through deep-sky surveys \citep{Getman2008, Pye2015, Kuznetsov2021, Zhao2024}. However, their narrow fields of view and typically short pointing durations substantially limit the detection efficiency, making them more effective for probing faint and short‑lived flares than for systematically discovering extreme events.
Consequently, it has remained challenging to obtain statistically robust constraints on extreme stellar flare occurrence rates, energy distributions, and the underlying physical mechanisms.

Successfully launched on 9 January 2024, the Einstein Probe (\ep; \citealt{Yuan2022}) provides a novel solution to overcome this bottleneck. \ep\ is a soft X-ray space observatory dedicated to the discovery and monitoring of a wide range of transient phenomena, with stellar flares among its primary scientific targets \citep{Yuan2025}. The mission carries two X-ray telescopes: the Wide-field X-ray Telescope (WXT; \citealt{Cheng2025}), which features a large field of view and an onboard real-time triggering system, and the Follow-up X-ray Telescope (FXT; \citealt{Chen2021}), designed to perform rapid, high-sensitivity follow-up observations. WXT data are processed onboard in real time to identify transient events or bursts exceeding predefined thresholds; the detection of such an event generally triggers an automatic FXT follow-up, with the spacecraft slewing to the target within approximately 3--5 minutes. In addition to these autonomous observations, FXT also supports Target of Opportunity (ToO) observations via command uplink. The key specifications of both instruments are summarized in Table~\ref{tab:instru}.

\ep-WXT employs novel lobster-eye micro-pore optics (MPO) coupled with an array of 48 back-illuminated scientific Complementary Metal-Oxide-Semiconductor (CMOS) detectors \citep{Wu2022, Wang2022}. This optical design was previously validated on orbit by the pathfinder mission Lobster-Eye Imager for Astronomy (\leia; \citealt{Zhang2022, Ling2023}), which detected the most energetic stellar X-ray flare reported to date from the star HD~251108 \citep{Ling2022, Mao2025}. Operating in the 0.5--4~keV energy band, \ep-WXT provides an instantaneous field of view of 3850~deg$^{2}$, corresponding to approximately one-eleventh of the entire sky, with an angular resolution of $\sim5\arcmin$ (full width at half maximum, FWHM) and a temporal resolution of 50~ms. The point-spread function (PSF) on the focal sphere exhibits a characteristic cruciform morphology with a bright central core (Figure~\ref{fig:image}). For a typical exposure of 1000~s, \ep-WXT reaches a sensitivity of $\sim(2$--$3)\times10^{-11}$~erg~cm$^{-2}$~s$^{-1}$, outperforming comparable wide-field X-ray monitors by 1--2 orders of magnitude. These instrumental and performance characteristics make \ep-WXT particularly effective for detecting rare, energetic, and long-duration stellar flares, as well as for capturing their early-time evolution and temporal substructure along with the rapid FXT follow-up observations.

\ep\ operates in a near-Earth orbit at an altitude of 592~km with an orbital period of 96~min. In survey mode, \ep-WXT typically scans the full night sky (i.e., the hemisphere opposite the Sun) every $\sim5$~hr, providing high-cadence, long-term, and largely unbiased monitoring of stellar X-ray activity. This observing strategy enables the systematic discovery of luminous stellar flares across a wide range of stellar types and evolutionary stages, and facilitates population-level studies of flare energetics, durations, and recurrence rates. By July 2025, \ep\ had completed its first year of scientific operations, yielding a substantial and rapidly growing dataset well suited for stellar time-domain studies; several notable stellar flare events have already been reported by \citet{Wang2024, Wang2026} and \citet{Zhao2026}.

In this paper, we present a pilot study of long-term monitoring of flaring stars by \ep\ focusing on the nearby RS CVn-type system $\sigma$ Gem. RS CVn stars are a class of highly active stars, typically consisting of a rapidly rotating late-type giant (or subgiant) and a cool main-sequence companion \citep{Hall1976}. Strong tidal locking allows these systems to maintain high activity even as they age, since the rapid rotation, sustained by the orbital angular momentum rather than by the rotational angular momentum of a single star, drives the magnetic activity. Therefore, age is not a reliable activity proxy in these systems, and such high levels of magnetic activity result in frequent and intense X-ray flares \citep{Pye1983, Osten1999, Sasaki2021, Karmakar2023}. Consequently, RS CVn binaries serve as ideal laboratories for studying the observational properties and statistical behavior of stellar flares.

\begin{table}[h!]
\caption{Basic Properties of $\sigma$ Gem.}
\label{tab:prop}
\begin{center}
\begin{tabular}{ccc}
\hline\hline
Property & Value & Reference \\
\hline
R.A. (degree) & 115.8283 & [1] \\
Dec (degree) & +28.8835 & [1] \\
Distance (pc) & 36.9$\pm$0.5 & [2] \\
Period (day) & $19.6027\pm0.0005$ & [3] \\
Age (Gyr) & 5$\pm$1 & [3] \\
G-band (mag) & 3.968$\pm$0.006 & [1] \\
$F_\mathrm{X}$\textsuperscript{$\dagger$} (erg cm$^{-2}$ s$^{-1}$) & $1.20\times10^{-10}$ & [4] \\
\hline
\multicolumn{3}{c}{Primary} \\
\hline
Spectral type & K1III & [5] \\
$T_\mathrm{eff}$ (K) & $4530\pm60$ & [3] \\
log [$g$ (cm s$^{-2}$)] & $2.54\pm0.02$ & [3] \\
$M_*$ ($M_\odot$) & $1.28\pm0.07$ & [3] \\
$R_*$ ($R_\odot$) & $10.1\pm0.04$ & [3] \\
\hline
\end{tabular}
\end{center}
\tablefoot{References: [1] \citet{Gaia2023}; [2] \citet{BailerJones2021}; [3] \citet{Roettenbacher2015}; [4] \citet{Freund2024}; [5] \citet{Duemmler1997}.}
\textsuperscript{$\dagger$} The {\it eROSITA} flux in 0.2--2.3 keV.
\end{table}

$\sigma$ Gem is a binary system comprising a K1-type red giant and an unseen companion, which is likely a late-type main-sequence star. Its basic parameters are listed in Table \ref{tab:prop}. As a bright X-ray source, $\sigma$ Gem has been detected by multiple X-ray missions \citep{Boller2016, Freund2024}, exhibiting a characteristic luminosity of $\sim10^{31}$ erg s$^{-1}$. Several X-ray flares from this star have been reported in the literature \citep{Pye1983, Sanz2002, Gudel2002, Nordon2006, Pandey2012, Serino2014}. With its high flux level (about an order of magnitude above the \ep-WXT detection threshold) and well-documented high activity, $\sigma$ Gem presents an ideal target for the first systematic analysis of stellar flares using \ep\ data.

This paper presents flare analysis of $\sigma$ Gem, based on the first-year observational data from the \ep-WXT. The paper is structured as follows. In Section \ref{sec:obs_data} we describe the \ep-WXT observations and the data reduction procedures. In Section \ref{sec:data_anal} we detail the temporal and spectral analysis of the flares identified by Bayesian Blocks method. Then, in Section \ref{sec:dis} we discuss the coronal and flaring properties, the flare frequency, and methodological limitations. Finally, in Section \ref{sec:conclu} we summarize the work and outline prospects for future research.

\section{Observations and data reduction}
\label{sec:obs_data}
After a series of in-orbit calibrations, \ep\ officially began its scientific operation in July 2024. 
During its first year of operation, $\sigma$ Gem was observable within a window of about six months, from October 18, 2024, to April 13, 2025, in which the star satisfied the Sun avoidance angle constraint for \ep-WXT. Over this period, the star fell within the FoV for 1055 observational pointings (each assigned a unique observation ID, or ObsID) and was detected a total of 1135 times by various CMOS detectors\footnote{The source could be detected simultaneously by two CMOS detectors if it fell within their overlapping FoV (e.g., between CMOS14 and CMOS37, CMOS15 and CMOS38, etc.) during a given pointing. This means $\sigma$ Gem fell in the overlapping FoVs in 80 of these 1055 pointings.}.

Additionally, \ep-FXT also conducted 9 ToO observations of $\sigma$ Gem during this one-year period. The data reduction and analysis of FXT observations are presented in Appendix \ref{sec:fxt}.

We then performed a screening of the 1135 detections by \ep-WXT. First, we excluded detections where the source fell within 9 arcmin of the detector FoV edge, as the auxiliary response file (ARF) was calibrated for a 9-arcmin aperture and spectra extracted from a different-sized region would introduce systematic errors in the flux and spectral analysis. We also removed detections located in known bad regions of CMOS41 and CMOS34, which exhibit anomalously low count rates. Following this screening process, a total of 1049 detections (from 986 ObsIDs) were retained for further processing and analysis, with a cumulative exposure time of $1.69\times10^6$ s.

\begin{figure}[h!]
\centering
\includegraphics[width=8cm]{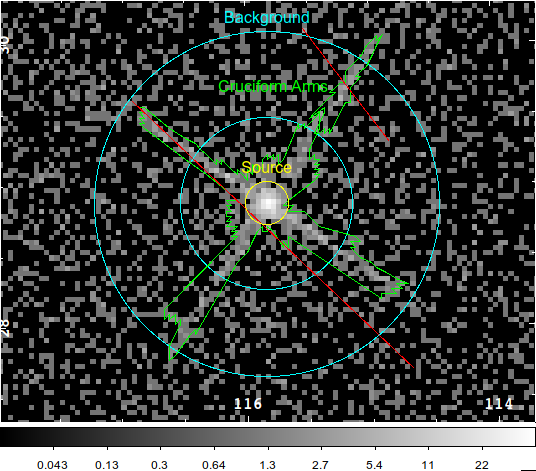}
\caption{An example \ep-WXT detection image of $\sigma$ Gem (ObsID: 11916647171, CMOS ID: 35). Source and background events are extracted from the yellow circular region and the cyan annular region, respectively. The cruciform arms (green regions), which are characteristic of the lobster‑eye MPO optics PSF,  are excluded from the background extraction.}
     \label{fig:image}
\end{figure}

\begin{figure*}[h!]
\centering
\includegraphics[width=15cm]{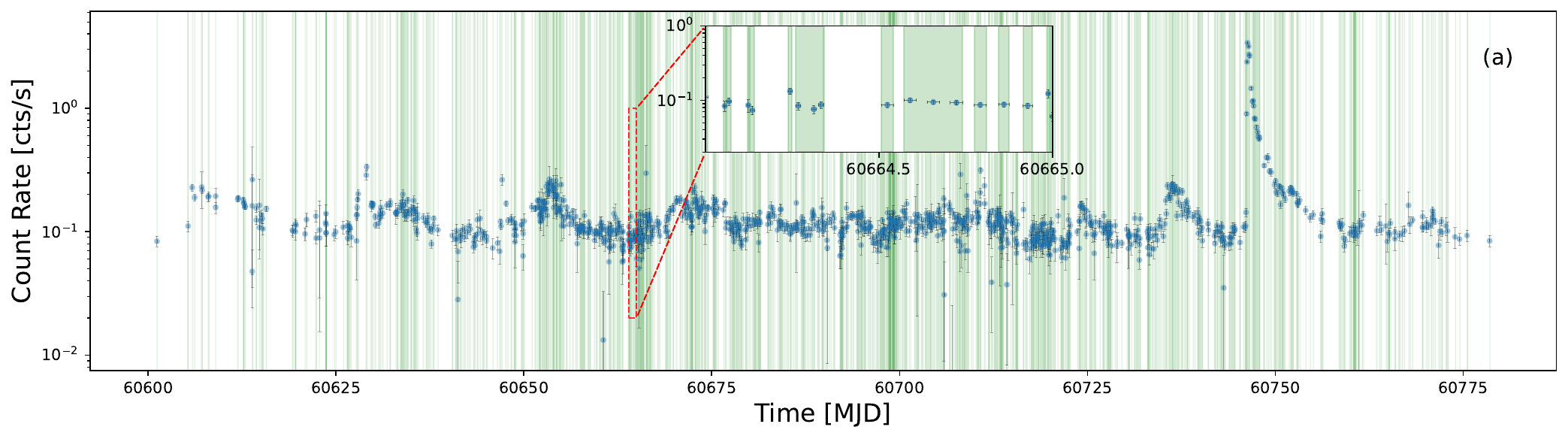}
\includegraphics[width=15cm]{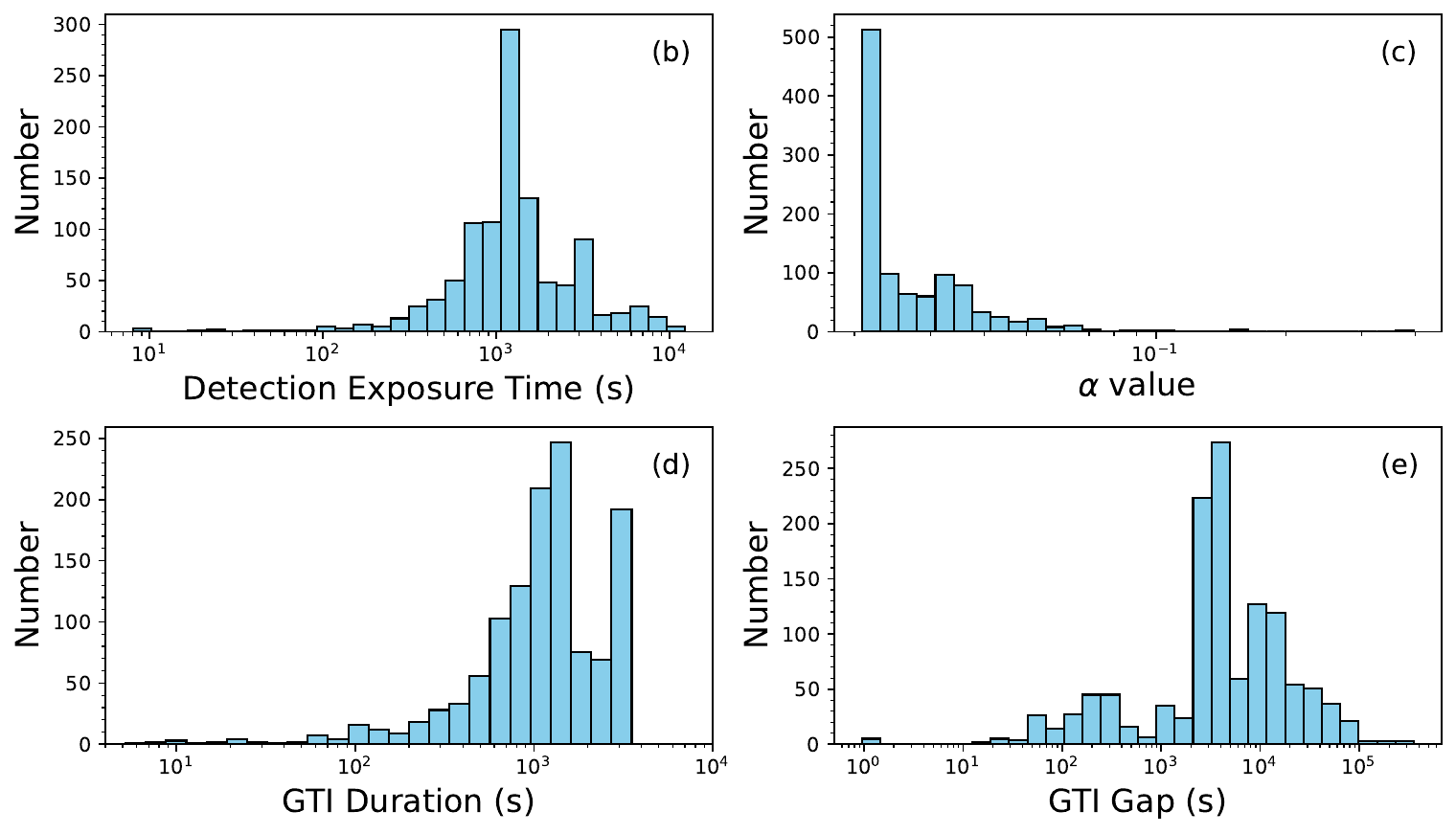}
\caption{(a) Light curve and (b)-(e) statistics of the 1049 \ep-WXT detections for $\sigma$ Gem. Each point in the light curve represents a single Good Time Interval (GTI), with the count rate and its uncertainty calculated by Equation (\ref{eq:rate})-(\ref{eq:rateerr}). The green bands mark the 1049 individual detections. A zoom-in view of the light curve enclosed by the red box (MJD 60664–60665) is shown in the inset to better illustrate the correspondence between the detections and the GTIs. (b) Distribution of detection exposure time. Each detection may consist of one or more GTIs. (c) Distribution of $\alpha$ values, defined as the ratio of areas between the source and background extraction regions. (d) Distribution of GTI duration. (e) Distribution of GTI gaps.
} 
     \label{fig:obs_stat}
\end{figure*}

The data reduction and calibration of X-ray events were performed with dedicated software and a calibration database (CALDB) for WXT (Liu et al., in prep.). The CALDB was constructed from on-ground calibration experiments \citep{Cheng2025}, a methodology previously validated on a WXT prototype instrument. For each detection, the 0.5--4 keV source and background events were extracted from a circular region with a 9-arcmin radius and an annular region with inner and outer radii of 36' and 72', respectively. The region of cruciform arms (the extended structure of the PSF), produced by the WXT data analysis software (WXTDAS), was excluded from the background region during extraction (see Figure \ref{fig:image}). The corresponding source and background spectra along with the response (RMF) and auxiliary response (ARF) files were also generated by WXTDAS, and the source-to-background area ratio $\alpha$ was calculated\footnote{The shape of cruciform arms differs for each detection, leading to variations in the arm-excluded background region. As a result, the $\alpha$ value varies from one detection to another.}.
The light curve and data sampling statistics are presented in Figure \ref{fig:obs_stat}. Since the Earth occultation interrupts observations lasting more than one orbit, the exposure time for an ObsID may be divided into one or more Good Time Intervals (GTIs). 

From the 1049 detections, we compiled combined source and background event lists containing 225318 and 765826 events\footnote{To be clear, an event is a single photon recorded by the detector. A detection, by contrast, requires a significant excess of source events over background events (i.e., a sufficient signal-to-noise ratio, SNR) in a given pointing.}, respectively. Both lists recorded, for each event, the arrival time, pixel position, energy channel, ObsID, CMOS ID, vignetting factor (which is determined by the photon's energy and its position on the detector) and an overlap-FoV flag\footnote{$\sim$12\% of events fell within the overlapping FoVs, consisting of 26430 source events and 89052 background events.} (true if the event falls within the region shared by two CMOS detectors). All GTIs were also included. The background list additionally included the scaling parameter $\alpha$ for each event, serving as a weight for the background count.

\section{Data analysis} \label{sec:data_anal}
\subsection{Bayesian blocks}
\label{sec:bb}

Bayesian Blocks is a method to identify change points in a time series, and segment it into blocks. It is widely used to identify and characterize flaring behavior in X-ray data \citep{Scargle2013, Abdo2013, Mossoux2015, Getman2021F, Zhao2024}. In this work, we used the function \texttt{\detokenize{bayesian_blocks}} from Astropy, a community-developed core Python package for Astronomy \citep{Astropy2013}, to fit the light curve with Bayesian Blocks.

\subsubsection{Light curve construction}
\label{sec: pre_work}


Applying Bayesian Blocks analysis requires the light curve to be binned into intervals of equal effective exposure time. To facilitate this, we constructed a new, continuous time series by concatenating all GTIs, eliminating gaps and setting the start time to zero. This continuous time series was then used for the subsequent analysis.

\begin{figure*}[h!]
\centering
\includegraphics[width=15cm]{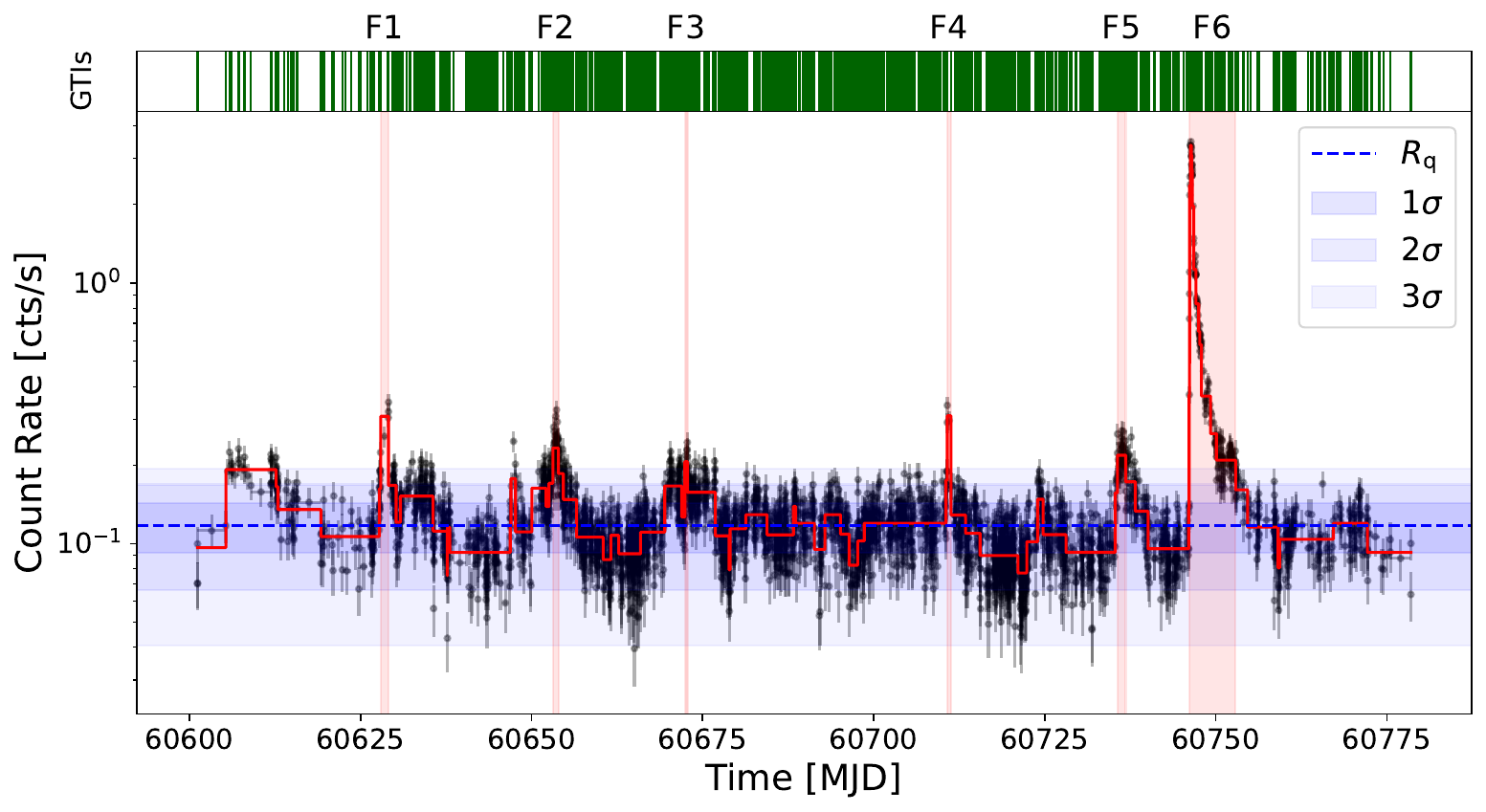}
\caption{The background-subtracted vignetting-corrected light curve constructed in Section \ref{sec: pre_work}. Each bin is shown as a black point, for which the effective exposure time (accumulated GTIs) is 500 s, except for the final point. Note that this binning scheme differs from that in Figure \ref{fig:obs_stat}, where each point represents a single GTI. The upper horizontal bars indicate good time intervals (GTIs) of the observations. The Bayesian blocks are represented by a red step function. The derived pseudo-quiescent count rate $R_\mathrm{q}$ is indicated by the blue dashed line. The light and dark blue shaded bands mark the 1--3 $\sigma$ range around $R_\mathrm{q}$, and flare blocks are highlighted in red.}
     \label{fig:bblock}
\end{figure*}

Furthermore, the time series was binned into a total of $N$ bins to obtain the background-subtracted vignetting-corrected count rate. The bin width $\Delta t$ was iteratively determined to ensure a minimum signal-to-noise ratio (SNR) of 3 per bin, which was calculated as $R_{i,\mathrm{net}}/\sigma(R_{i,\mathrm{net}})$. Here, $R_{i,\mathrm{net}}$ and $\sigma(R_{i,\mathrm{net}})$ are the net count rate and its uncertainty in the $i$-th bin, respectively.
Starting from an initial value of 100 s, $\Delta t$ was increased in steps of 100 s until this criterion was met. The last bin of the time series, if shorter than $\Delta t$, was merged with the preceding bin. A final bin width of $\Delta t = 500$s was adopted, dividing the light curve into $N=3372$ bins (black points in Figure \ref{fig:bblock}). 

The net count rate in the $i$-th bin is calculated as:
\begin{equation}
\label{eq:rate}
\begin{aligned}
R_{i,\mathrm{net}} &= \frac{c_{i,\mathrm{src}}-c_{i,\mathrm{bkg}}}{\Delta t_i} \\
&=\frac{\sum_{j}{(f_{i,j}/v_{i,j})}-\sum_{k}{(f_{i,k}\alpha_{i,k}/v_{i,k})}}{\Delta t_i}, i=1,2,...N
\end{aligned}
\end{equation}
Here, $c_{i,\mathrm{src}}$ and $c_{i,\mathrm{bkg}}$ are the effective source and background counts in the $i$-th bin. $f_{i,j}$ and $f_{i,k}$ are the FoV‑overlap factors for the $j$-th source event and the $k$-th background event in the $i$-th time bin, respectively. The factor is 0.5 if the event lies within the overlapping FoV of two CMOS detectors, and 1 otherwise.
Similarly, $v_{i,j}$ and $v_{i,k}$ denote the corresponding vignetting factors, and $\alpha_{i,k}$ is the corresponding background scaling parameter. The time bin width is fixed at $\Delta t_i = 500$ s for all bins except the last one ($\Delta t_N$).

The associated uncertainty is estimated as:
\begin{equation}
\label{eq:rateerr}
\begin{aligned}
\sigma(R_{i,\mathrm{net}}) &= \frac{\sqrt{\sigma^2(c_{i,\mathrm{src}})+\sigma^2(c_{i,\mathrm{bkg}})}}{\Delta t_i} \\
&= \frac{\sqrt{\sum_{j}{(f_{i,j}/v_{i,j})}+\sum_{k}{(f_{i,k}\alpha_{i,k}/v_{i,k})}}}{\Delta t_i}, i=1,2,...N
\end{aligned}
\end{equation}
Here, the uncertainty for the counts $c$ in the $i$-th bin is calculated as $\sigma(c_i)=\sqrt{c_i}$, following Poisson statistics. 

\subsubsection{False Positive Rate}
\label{sec: fpr}
When applying Bayesian Blocks, a parameter named \texttt{\detokenize{ncp_prior}} is usually required. This parameter is the prior on the number of change points, which controls the sensitivity of change-point detection. The value of \texttt{\detokenize{ncp_prior}} depends only on the number of data points $N$ and the false positive rate $p_0$ (i.e., the probability of falsely reporting detection of a change point), according to \citet{Scargle2013}
$$
\mathrm{ncp_{prior}}=\psi(N,p_0)
$$

\begin{table*}[h!]
\centering
\caption{Best-fit Spectral Parameters for Pseudo-quiescent State Observations.}
\label{tab:qui_fit}
\begin{tabular}{ccccc}
\hline\hline
Parameters & 1-T APEC & 2-T APEC & 3-T APEC & 4-T APEC \\
\hline
$N_\mathrm{H}$ (10$^{21}$ cm$^{-2}$)                 & 1.03       & 1.64$_{-0.07}^{+0.08}$    & 1.01$_{-0.01}^{+0.01}$    & 1.19$_{-0.01}^{+0.02}$    \\
$Z$ ($Z_{\odot}$)                         & 0.053       & 0.079$_{-0.003}^{+0.003}$ & 0.17$_{-0.02}^{+0.02}$ & 0.18$_{-0.02}^{+0.02}$ \\
$kT_\mathrm{q,1}$ (keV)                   & 1.03       & 0.28$_{-0.01}^{+0.01}$ & 0.30$_{-0.01}^{+0.02}$ & 0.21$_{-0.04}^{+0.03}$ \\
EM$_\mathrm{q,1}$ ($10^{54}$ cm$^{-3}$)   & 3.54       & 2.0$_{-0.2}^{+0.3}$   & 0.54$_{-0.09}^{+0.11}$   & 0.47$_{-0.18}^{+0.18}$   \\
$kT_\mathrm{q,2}$ (keV)                   & -          & 1.10$_{-0.01}^{+0.01}$    & 1.02$_{-0.01}^{+0.02}$    & 0.41$_{-0.08}^{+0.10}$    \\
EM$_\mathrm{q,2}$ ($10^{54}$ cm$^{-3}$) & -          & 3.2$_{-0.1}^{+0.1}$    & 1.1$_{-0.1}^{+0.1}$    & 0.32$_{-0.13}^{+0.28}$    \\
$kT_\mathrm{q,3}$ (keV)                   & -          & -                      & 2.12$_{-0.10}^{+0.13}$ & 1.03$_{-0.01}^{+0.02}$ \\
EM$_\mathrm{q,3}$ ($10^{54}$ cm$^{-3}$) & -          & -                      & 1.1$_{-0.1}^{+0.1}$    & 1.1$_{-0.1}^{+0.2}$    \\
$kT_\mathrm{q,4}$ (keV)                   & -          & -                      & - & 2.17$_{-0.12}^{+0.17}$ \\
EM$_\mathrm{q,4}$ ($10^{54}$ cm$^{-3}$) & -          & -                      & -    & 1.1$_{-0.1}^{+0.1}$    \\
$\chi^2_\nu$ (d.o.f.)                    & 2.42 (329) & 1.62 (327)             & 1.18 (325)             & 1.17 (323)            
\\
\hline
\end{tabular}
\tablefoot{All errors represent the $1 \sigma$ uncertainties. $N_\mathrm{H}$, $Z$, $kT$, and EM are the Galactic H$_\mathrm{I}$ column density, the metallicity, plasma temperature, and emission measures, respectively.}
\end{table*}

Therefore, for a given dataset with a fixed size 
$N$, the value of \texttt{\detokenize{ncp_prior}} is uniquely determined once $p_0$ is specified. In our case, $N$ was established in Section 3.1. To find the \texttt{\detokenize{ncp_prior}} corresponding to a false positive rate of $p_0 = 0.01$ following \citet{Scargle2013}, we conducted a series of simulations, which yielded an optimal value of \texttt{\detokenize{ncp_prior}} = 8.3 (see Appendix \ref{sec:ncp} for details).

\subsubsection{Flare identification}\label{sec:flare_id}

With the \texttt{\detokenize{ncp_prior}} derived in Section \ref{sec: fpr}, we applied the Bayesian Blocks analysis to the binned light curve from Section \ref{sec: pre_work}. This process detected 84 change points, which inherently include the start and end points of the light curve, thus dividing it into 83 blocks. Figure \ref{fig:bblock} presents the segmented light curve on the true time axis, where each data point has an effective exposure time of 500 s except for the final one.

We then calculated the weighted mean count rate, $R_\mathrm{q}$, which was derived by weighting the block count rates by their respective durations. Since the block durations are unequal, we adopted the weighted mean instead of the median, as it properly weights the time spent at each count rate level.\footnote{We also calculated the median count rate of the individual data points (0.116 counts/s), which is comparable to the weighted mean of the blocks (0.117 counts/s), confirming that the choice of the weighted mean does not bias the result.} Blocks with count rates exceeding $R_\mathrm{q}$ by more than $3\sigma(R_\mathrm{q})$ (where $\sigma(R_\mathrm{q})$ is the uncertainty of $R_\mathrm{q}$ ) were iteratively removed, and $R_\mathrm{q}$ was recalculated until no such outliers remained. We define blocks with count rates within $1\sigma$ of the final $R_\mathrm{q}$ ($0.117\pm0.026$ counts/s) as the "pseudo-quiescence", and $R_\mathrm{q}$ itself is designated as the "pseudo-quiescent" count rate. Here, the term "pseudo-quiescence" is used to emphasize that the observed intervals with count rates within $R_\mathrm{q}\pm3\sigma(R_\mathrm{q})$ are not devoid of activity. Rather, it indicates that the dominant stellar activity falls below our instrumental detection threshold. A further discussion of the activity during pseudo‑quiescence in Section \ref{sec:coro_pro} should help to better understand this term. In this work, the pseudo-quiescent spectrum serves as a practical baseline for investigating the properties of the flaring plasma. Based on this definition, one or more adjoining blocks with count rates above $R_\mathrm{q}+3\sigma(R_\mathrm{q})$ were identified as a flare. In total, 6 flares were identified through this process (Figure \ref{fig:bblock}) and are designated chronologically as F1-F6.

\subsection{Spectral analysis}
\label{sec:data_ana}

To investigate the spectral properties and evolution of the star, we performed X-ray spectral fitting employing the Xspec package \citep{Arnaud1996}. Specifically, we adopted the thin thermal plasma model, APEC \citep{Smith2001}, with one or multiple temperature components. TBabs \citep{Wilms2000} was used for the absorption. The required spectra were obtained by merging individual‑detection spectra (generated by WXTDAS) from the corresponding time intervals using the task \texttt{mathpha}, weighted by exposure time. The background spectra were further scaled by the source‑to‑background area ratio $\alpha$ before combination. Corresponding response files were combined with the task \texttt{ftaddrmf}, also weighted by exposure time.

\subsubsection{Pseudo-quiescent state} \label{sec:qui}

The pseudo‑quiescent spectrum was constructed from all photon events outside the flare phases. Initially, we tentatively fitted the spectra using 1 to 4 APEC thermal plasma components (specifically, \texttt{tbabs*apec}, \texttt{tbabs*(apec+apec)}, \texttt{tbabs*(apec+apec+apec)}, and \texttt{tbabs*(apec+apec+apec+apec)}), with the redshift fixed at zero. For models with multiple APEC components, the metallicity ($Z$) was set to be the same across all components, while the remaining parameters were left free. The best-fit parameters of the pseudo-quiescent spectrum are listed in Table \ref{tab:qui_fit}.

\begin{figure}[h!]
\centering
\includegraphics[width=\hsize]{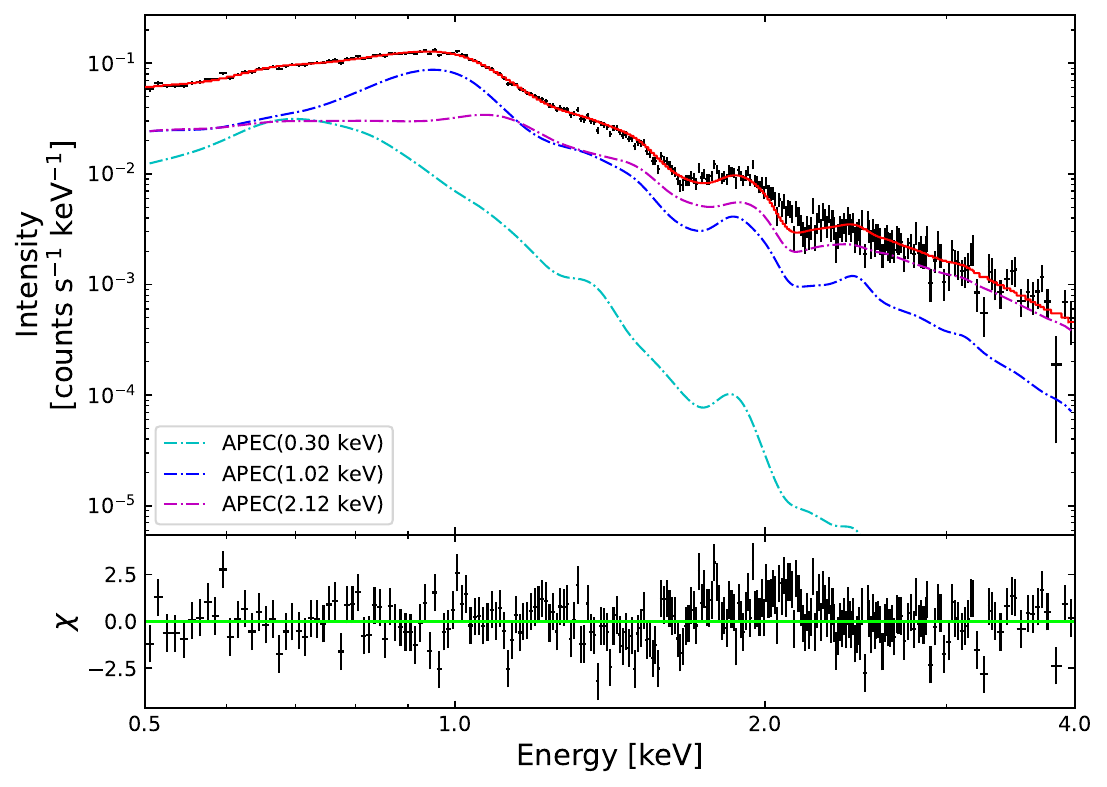}
\caption{Pseudo-quiescent X-ray spectrum, fitted by 3-T APEC model (red solid line) composed of  three components (dashdot).}
     \label{fig:qui_spec}
\end{figure}

By comparing these four models using an F-test, we found that a three-temperature (3-T) APEC model best reproduces the pseudo-quiescent spectrum (Figure \ref{fig:qui_spec}). Adding a fourth component did not significantly improve the fitting (here and subsequently, "significant" denotes an improvement exceeding the $3\sigma$ confidence level).

Furthermore, we applied the \texttt{cflux} model to derive the unabsorbed flux in the 0.5--4 keV band, which yielded a flux of $F_\mathrm{X,q}$ = $1.2\times10^{-10}$ erg cm$^{-2}$ s$^{-1}$ and a corresponding luminosity of $L_\mathrm{X,q}$ = $2.0\times10^{31}$ erg s$^{-1}$.

\begin{figure*}[h!]
\centering
\includegraphics[width=\hsize]{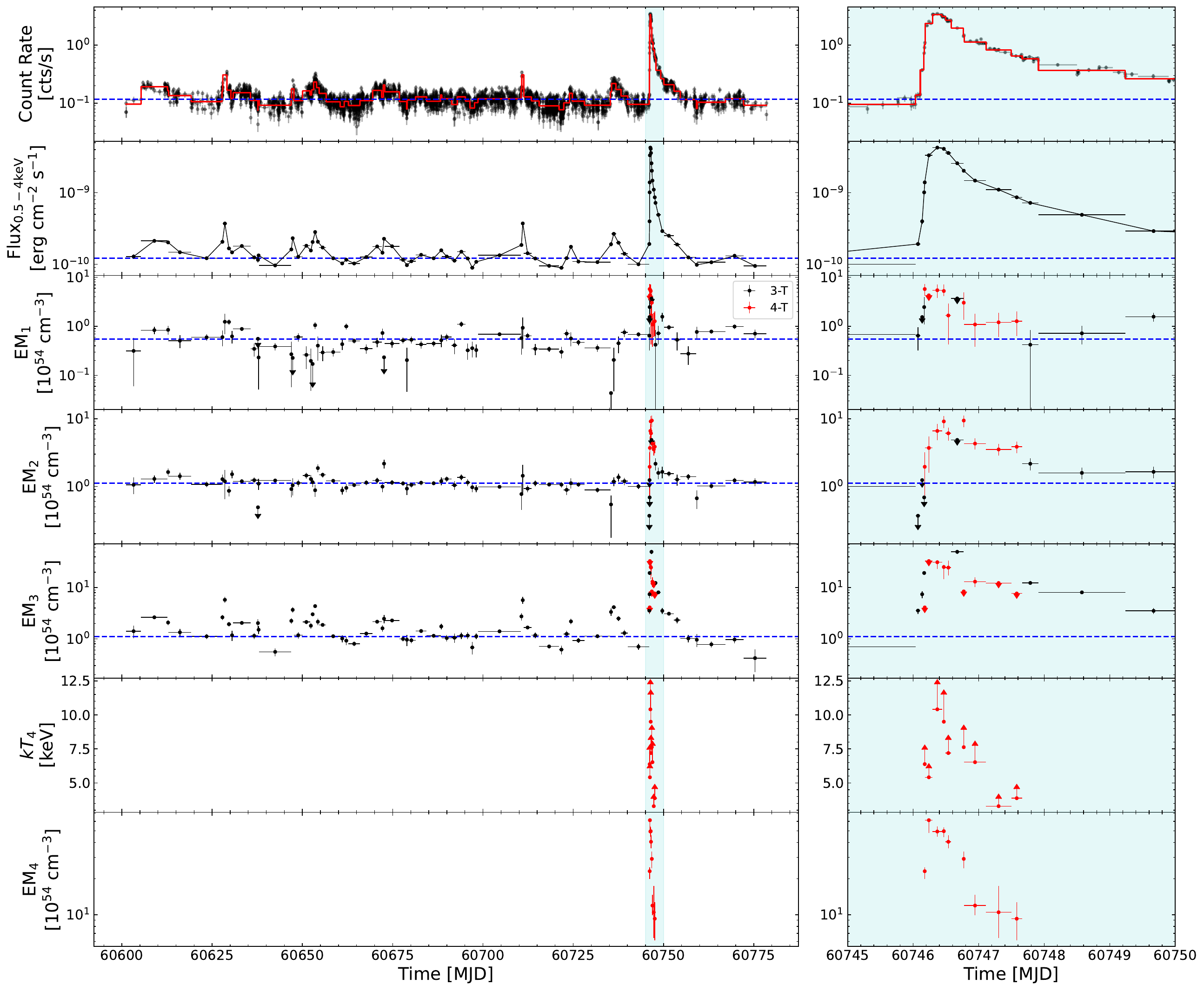}
\caption{X-ray light curve and temporal evolution of spectral parameters for $\sigma$ Gem. 
From top to bottom, the panels show the vignetting-corrected net count-rate light curve in 0.5--4 keV (identical to Figure \ref{fig:bblock}), absorption-corrected flux in the same band, time evolution of EMs for the three pseudo-quiescent plasma components (with temperatures fixed to their pseudo-quiescent values), time evolution of the temperature and EM of the fourth component. In the bottom five panels, each point corresponds to the spectral fit result per Bayesian block. Black points indicate the 3‑T APEC fit results, and red points show the 4‑T APEC fit results. Blue dashed lines in top four panels indicate the best-fit pseudo-quiescent values (see Table \ref{tab:qui_fit}).
The right panels are the zoom-in views of the cyan-shaded time intervals in the left panels, focusing on F6 to highlight the detailed evolution. Similar zoom-ins for F1-F5 are provided in Figure \ref{fig:spec_evo2}.}
     \label{fig:spec_evo}
\end{figure*}

As noted earlier, the pseudo-quiescent spectrum discussed here represents the average spectrum during the star’s relatively inactive periods. A good fit with a 3‑T APEC model does not imply that the spectrum originates from three distinct physical components. Rather, the model suggests that the spectral shape can be adequately represented by three temperature components, which are best interpreted as effective temperatures.

\subsubsection{Block‑by‑block analysis}
\label{sec:bw_ana}
Based on the best-fit parameters derived from the pseudo-quiescent spectrum, we performed block-by-block spectral fitting using both three- and four-component APEC models. We fixed the following parameters to their pseudo-quiescent values: the hydrogen column density ($N_\mathrm{H}$), the tied metallicity ($Z$), and the temperatures of the first three APEC components (redshift was fixed at zero for all). Meanwhile, all emission measures (EM) were left free. For the 4-T APEC model, the fourth temperature was also free to vary.

An F-test confirmed that the spectra of most blocks were well-reproduced by the 3-T APEC model. Only seven blocks during F6 (red data points in Figure \ref{fig:spec_evo}) showed a preference for a 4-T APEC fit, and introducing a fifth component yielded no significant improvement to the fit. Consequently, we adopted the 4-T APEC model for these seven blocks and the 3-T APEC model for all remaining blocks. The temporal evolution of the relevant parameters is presented in Figure \ref{fig:spec_evo}.

\subsection{Flare analysis}

\begin{table*}[h!]
\caption{Best-fit Parameters of the X-ray Light Curve}
\label{tab:time_para}
\centering
\begin{tabular}{cccccccc}
\hline\hline
Flare & $t_\mathrm{p}$\textsuperscript{a} & $R_\mathrm{p}$\textsuperscript{b} & $R_\mathrm{q,1}$\textsuperscript{c} & $R_\mathrm{q,2}$\textsuperscript{d} & $\tau_\mathrm{r}$\textsuperscript{e} & $\tau_\mathrm{d}$\textsuperscript{f} &
$\chi^2_\nu$ (dof)\textsuperscript{g} \\
 & (MJD) & (counts s$^{-1}$) & (counts s$^{-1}$) & (counts s$^{-1}$) & (ks) & (ks) & \\
\hline
F1 & $60629.091\pm0.003$ & $0.34\pm0.03$ & $0.104\pm0.003$ & $0.109\pm0.007$ & $139.4\pm3.5$ & $40.6\pm4.1$ & 1.46(116)\\ 
F2 & $60653.642\pm0.003$ & $0.33\pm0.03$ & $0.152\pm0.003$ & $0.102\pm0.001$ & $49.0\pm2.7$ & $59.9\pm2.2$ & 2.33(338)\\ 
F3 & $60672.764\pm0.020$ & $0.24\pm0.02$ & $0.127\pm0.004$ & $0.113\pm0.001$ & $35.0\pm2.6$ & $48.6\pm3.2$ & 3.00(235)\\ 
F4 & $60710.764\pm0.003$ & $0.34\pm0.03$ & $0.112\pm0.001$ & $0.087\pm0.001$ & $3.7\pm0.4$ & $71.7\pm2.7$ & 2.03(721)\\
F5 & $60736.245\pm0.004$ & $0.27\pm0.02$ & $0.089\pm0.002$ & $0.093\pm0.002$ & $100.2\pm4.0$ & $130.7\pm6.8$ & 1.83(294)\\ 
F6 & $60746.368\pm0.061$ & $3.48\pm0.08$ & $0.094\pm0.002$ & $0.107\pm0.002$ & $22.9\pm0.1$ & $47.6\pm0.4$, $189.6\pm5.2$ & 4.01(309) \\ 
\hline
\end{tabular}
\tablefoot{All errors represent the 1 $\sigma$ uncertainty.\\
\textsuperscript{a}Flare peak time. \\
\textsuperscript{b}Flare peak count rate. \\
\textsuperscript{c}The mean count rate before flare.  \\
\textsuperscript{d}The mean count rate after flare. \\
\textsuperscript{e}Flare rise time ($t_\mathrm{p}-t_\mathrm{ST}$). \\
\textsuperscript{f}Flare e-folding decay time. \\
\textsuperscript{g}$\chi^2_\nu$ is the reduced $\chi^2$ and dof stands for degrees of freedom.
}
\end{table*}

\begin{figure*}[h!]
\centering
\includegraphics[width=15cm]{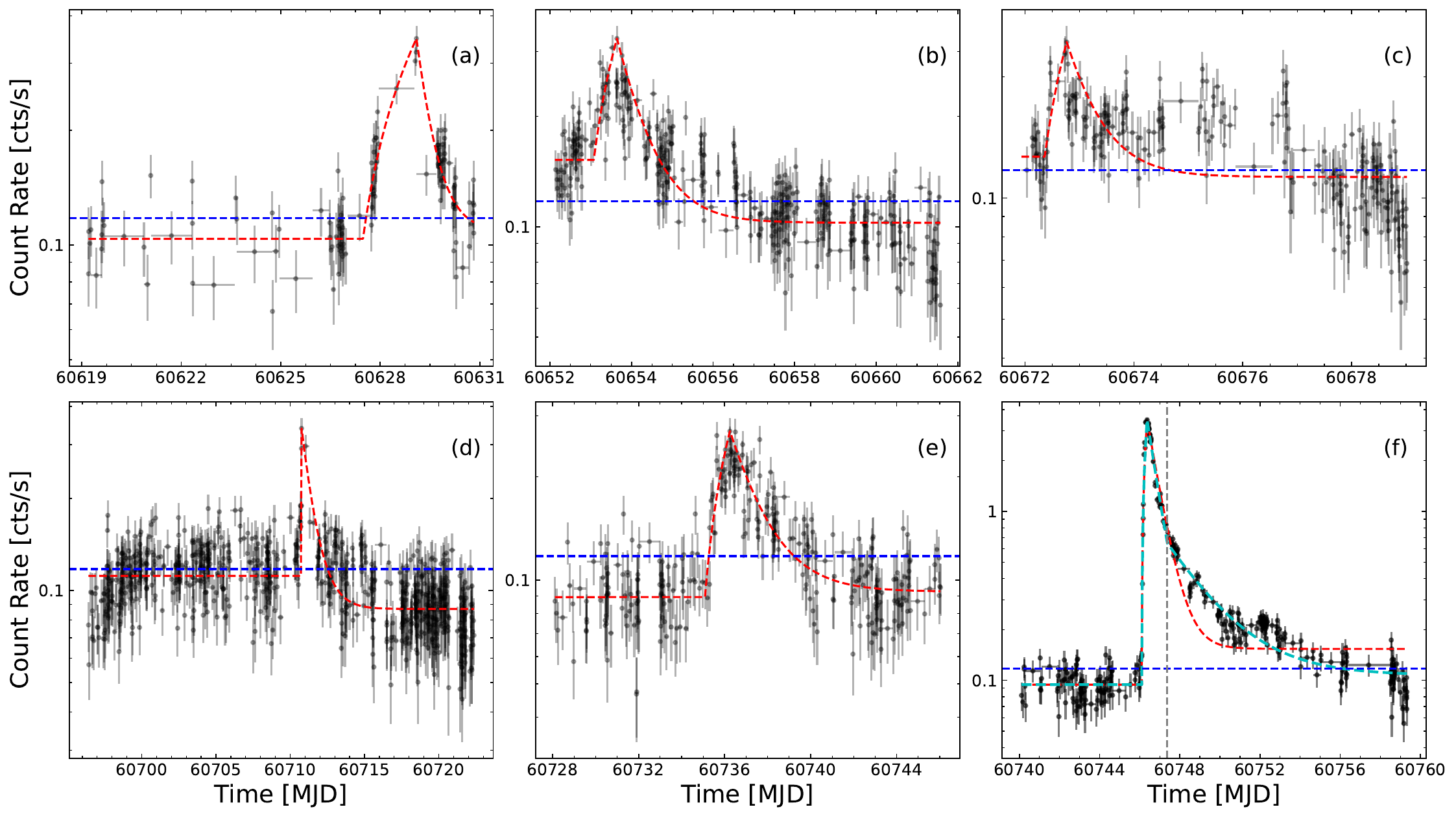}
\caption{(a)-(f) Light curves of flares F1-F6, respectively. The red dashed lines show the fits with a Fast Rise Exponential Decay (FRED) model. For F6, an alternative double-exponential model is overplotted in cyan in panel (f), with the vertical grey dashed line indicating the turning time point. The pseudo-quiescent count rate $R_\mathrm{q}$ is shown by the blue dashed lines.}\label{fig:lcurves}
\end{figure*}

For better characterization of the flares identified by Bayesian Blocks (Section \ref{sec:flare_id}), we fitted the binned light curve (black points in Figure \ref{fig:bblock}) with the Fast (Linear) Rise and Exponential Decay (FRED) model, a widely adopted model for stellar X-ray flares \citep{Reale2004, Wolk2005, Schrijver2012, Tsuboi2016, Sasaki2021}. Within this model, the count rate 
$R(t)$ at time $t$ is expressed as:
\begin{equation}
R(t) = \left\{
\begin{array}{lll}
R_\mathrm{q,1}& ,&{(t < t_\mathrm{ST})};\\
(R_\mathrm{p}-R_\mathrm{q,1})\times\frac{t-t_\mathrm{ST}}{t_\mathrm{p}-t_\mathrm{ST}}+R_\mathrm{q,1}& ,&{(t_\mathrm{ST} \leq t < t_\mathrm{p})};\\
(R_\mathrm{p}-R_\mathrm{q,2})e^{-(t-t_\mathrm{p})/\tau_\mathrm{d}}+R_\mathrm{q,2}& ,&{(t_\mathrm{p} \leq t).}
\end{array}
\right.
\end{equation}
Here, $R_\mathrm{q,1}$ and $R_\mathrm{q,2}$ are the mean count rates before and after the flare, respectively. $t_\mathrm{p}$ and $R_\mathrm{p}$ represent the flare peak time and the corresponding count rate. The rise timescale is defined as $\tau_\mathrm{r} = t_\mathrm{p} - t_\mathrm{ST}$, where $t_\mathrm{ST}$ is the flare start time, and $\tau_\mathrm{d}$ is the e-folding decay timescale.

For each flare, we selected a time interval to fit the light curve and derive the rise and decay timescales. To do this, we searched backward and forward from the flare peak to find the first local minimum block where count rate drops below $R_\mathrm{q}+\sigma(R_\mathrm{q})$ (i.e., within the $1\sigma$ uncertainty of the pseudo-quiescent level). The fitting interval then started at the beginning of the backward local minimum block and ended at the end of the forward local minimum block. During the fitting process, $t_\mathrm{p}$ and $R_\mathrm{p}$ were fixed to the observed flare peak time and count rate, respectively, while the remaining parameters were derived from the fitting (see Table \ref{tab:time_para}). The relatively large $\chi^2_\nu$ for some fits reflect the inherent simplicity of the FRED model, which cannot fully capture the complexity and scatter of the real light curves. Nevertheless, we consider this model sufficient for estimating the characteristic rise and decay timescales, which are the primary parameters of interest.

Following the initial fit, the decay phase of F6 exhibited a clear deviation from a single exponential model (red dashed line in Figure \ref{fig:lcurves}(f)). Specifically, it decayed more slowly in its later phase. This pattern in the light curve is commonly observed in both solar flares \citep{Aschwanden2001, Kashapova2021} and stellar flares \citep{Getman2008, Wargelin2008, Mao2025} and typically indicates a change of the dominant cooling process and the existence of sustained heating during the decay \citep{Reale2004, Reale2007}.

To account for this, we employed a double-exponential model to fit the decay phase. In this model, the light curve before and after the turning time point $t_\mathrm{T}$ is described by two distinct exponential decay phases:
\begin{equation}
R(t) = \left\{
\begin{array}{lll}
R_\mathrm{q,1}& , &  {(t < t_\mathrm{ST})};\\
(R_\mathrm{p}-R_\mathrm{q,1})\times\frac{t-t_\mathrm{ST}}{t_\mathrm{p}-t_\mathrm{ST}}+R_\mathrm{q,1}& , &  {(t_\mathrm{ST} \leq t < t_\mathrm{p})};\\
(R_\mathrm{p}-R_\mathrm{q,2})e^{-(t-t_\mathrm{p})/\tau_\mathrm{d,1}}+R_\mathrm{q,2}& , &  {(t_\mathrm{p} \leq t < t_\mathrm{T})};\\
(R_\mathrm{T}-R_\mathrm{q,2})e^{-(t-t_\mathrm{T})/\tau_\mathrm{d,2}}+R_\mathrm{q,2}& , &  {(t_\mathrm{T} \leq t)}.
\end{array}
\right.
\end{equation}
where $\tau_\mathrm{d,1}$ and $\tau_\mathrm{d,2}$ represent the e-folding decay timescales before and after $t_\mathrm{T}$, respectively, and $R_\mathrm{T}$ is the count rate at $t_\mathrm{T}$.

\begin{table*}[h!]
\caption{Summary of Derived Flare Properties}
\label{tab:energy_para}
\centering
\begin{tabular}{cccccccc}
\hline\hline
Flare & $\tau$\textsuperscript{a} & $F_\mathrm{X,p}$ & $f_{\mathrm{flare}}$\textsuperscript{b} & $L_\mathrm{X,p}$\textsuperscript{e} & $E_\mathrm{X}$\textsuperscript{f} \\
& (ks) & ($10^{-10}$ erg cm$^{-2}$ s$^{-1}$) & & ($10^{31}$ erg s$^{-1}$) & ($10^{36}$ erg) \\
\hline
F1 & $180.0\pm6.5$ & $3.71_{-0.19}^{+0.20}$ & $2.04_{-0.17}^{+0.18}$ & $6.05_{-0.35}^{+0.36}$ & $4.49_{-0.43}^{+0.44}$ \\ 
F2 & $109.0\pm5.3$ & $2.82\pm0.06$ & $1.32_{-0.07}^{+0.08}$ & $4.60_{-0.16}^{+0.15}$ & $2.21\pm0.15$ \\ 
F3 & $83.6\pm7.1$ & $2.27\pm0.11$ & $0.86_{-0.09}^{+0.10}$ & $3.70\pm0.20$ & $1.13_{-0.14}^{+0.15}$ \\ 
F4 & $75.3\pm3.9$ & $3.72_{-0.22}^{+0.23}$ & $2.05_{-0.19}^{+0.20}$ & $6.07_{-0.39}^{+0.40}$ & $3.00\pm0.31$ \\
F5 & $230.9\pm10.7$ & $2.66\pm0.06$ & $1.18\pm0.07$ & $4.34\pm0.15$ & $4.25\pm0.32$ \\ 
F6 & $260.0\pm10.4$ & $42.7_{-0.7}^{+0.4}$ & $34.03\pm0.95$ & $69.6_{-2.2}^{+1.9}$ & $43.5_{-1.4}^{+1.3}$ \\ 
\hline
\end{tabular}
\tablefoot{All errors represent the 1 $\sigma$ uncertainty.\\
\textsuperscript{a}Flare duration: $\tau=\tau_\mathrm{r}+\tau_\mathrm{d}$ for F1-F5 and $\tau=\tau_\mathrm{r}+\tau_\mathrm{d,1}+\tau_\mathrm{d,2}$ for F6.\\
\textsuperscript{b}The flare fractional flux variation $f_{\mathrm{flare}} = (F_\mathrm{X,p}-F_\mathrm{X,q})/F_\mathrm{X,q}$. \\
\textsuperscript{e}Flare peak luminosity in 0.5--4.0 keV: $L_\mathrm{X,p}=4\pi d^2F_\mathrm{X,p}$. \\
\textsuperscript{f}Total flare energy released: $E_\mathrm{X}=(L_\mathrm{X,p}-L_\mathrm{X,q})(\tau_\mathrm{r}/2+\tau_\mathrm{d})$ for F1-F5 and $E_\mathrm{X}=(L_\mathrm{X,p}-L_\mathrm{X,q})(\tau_\mathrm{r}/2+\tau_\mathrm{d,1})+(L_\mathrm{X,T}-L_\mathrm{X,q})(\tau_\mathrm{d,2}-\tau_\mathrm{d,1})$ for F6.
}
\end{table*}

Based on the fitting results, a notable feature is that for all flares except F1, the rise timescale is shorter than the decay timescale. Defining the total flare duration as $\tau = \tau_\mathrm{r} + \tau_\mathrm{d}$ (or $\tau = \tau_\mathrm{r} + \tau_\mathrm{d,1} + \tau_\mathrm{d,2}$ for F6), we found that these flares lasted from approximately 21 hours to about 3 days. These durations are notably longer than those of previously reported flares on the same star, which lasted for 12-13.5 hours, 15 hours and about 1 day (\citealt{Pye1983,Sanz2002,Gudel2002,Nordon2006,Nordon2008,Pandey2012}). This is largely because \ep’s long time baseline makes it more sensitive to longer-duration flares than previous instruments, rather than reflecting an increase in the star's activity level.

Meanwhile, the data show an increase in EM for all plasma components during flares, with the hotter components exhibiting a larger enhancement. This is exemplified by F6, where the EM increments for the four components (ordered by increasing temperature) are approximately 5$\times10^{54}$, 8$\times10^{54}$, 49$\times10^{54}$ and 61$\times10^{54}$ cm$^{-3}$ above their pseudo-quiescent values (or above zero for EM$_4$). This is consistent with the result of \citet{Nordon2006}, who reported from another X-ray flare on $\sigma$ Gem that the flaring activity had less influence on the cooler, global corona.

Due to the limited energy bandpass of \ep-WXT, the value of $kT_4$ cannot be well-constrained. However, it is still observable that its lower limit varies, showing a phase coherence with the rise and decay of the overall X-ray flux.

\section{Results and discussion}
\label{sec:dis}
\subsection{Coronal properties}
\label{sec:coro_pro}

\begin{figure}[h!]
\centering
\includegraphics[width=\hsize]{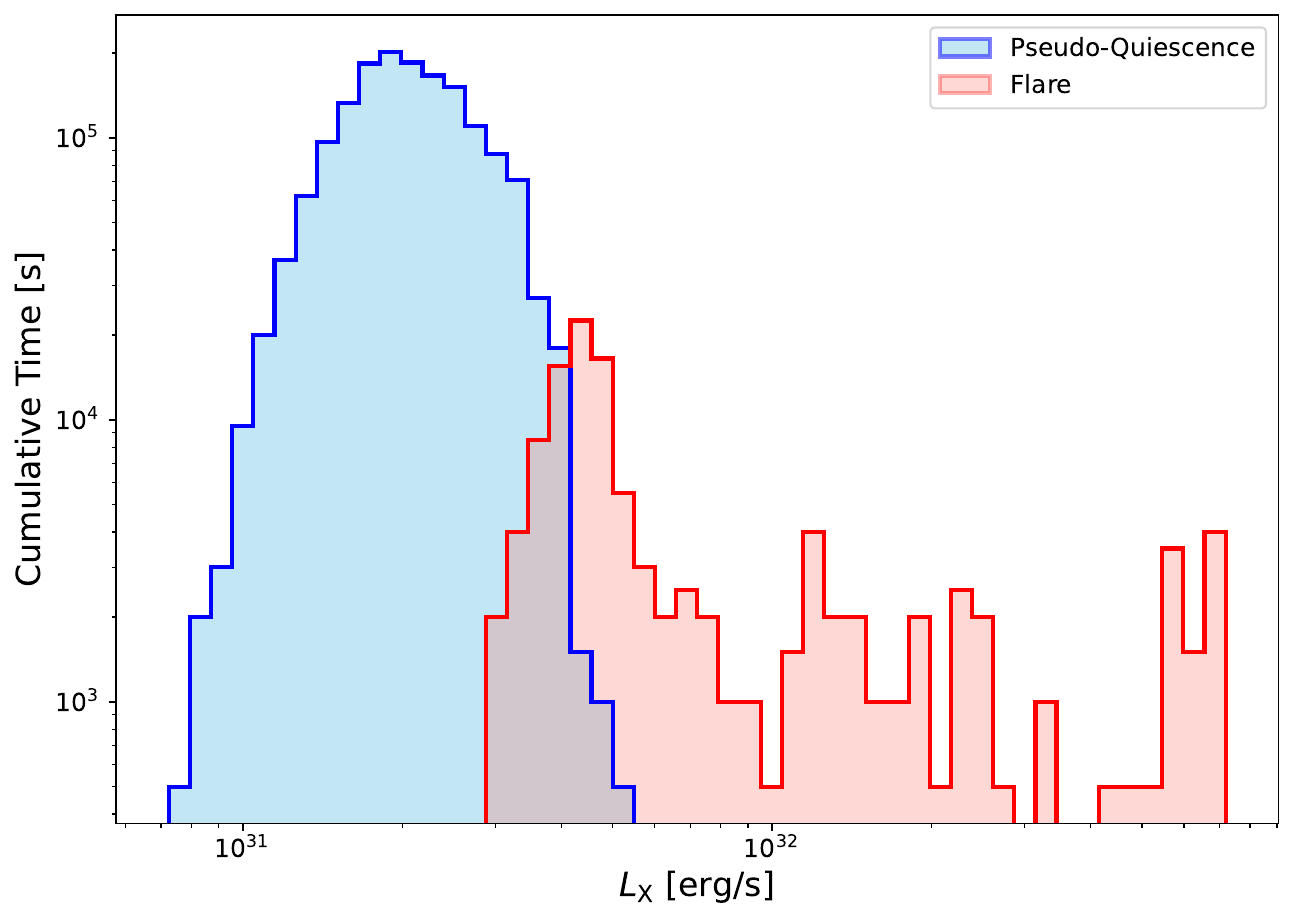}
\caption{Cumulative time spent at different X-ray luminosity levels for $\sigma$ Gem. The blue bars represent the time spent in the pseudo-quiescent state within each luminosity bin, while the red bars show the cumulative duration of flare activity. The luminosity is in 0.5--4 keV.}
     \label{fig:lum_dis}
\end{figure}

To analyze the coronal properties of $\sigma$ Gem, we performed a statistical analysis of the temporal distribution of the stellar X-ray luminosity (Figure \ref{fig:lum_dis}). Specifically, for each Bayesian block (see Section \ref{sec:bb}), the conversion factor is defined as the ratio between the luminosity derived from the block-by-block spectral fitting (see second panel of Figure \ref{fig:spec_evo} for the corresponding flux) and its mean count rate. This procedure took into account the instrumental response, and was necessary because the conversion from count rate to flux/luminosity is instrument dependent and also depends on the intrinsic spectral shape. Assuming that the spectrum remains constant within each block, the luminosity for each data point in Figure \ref{fig:bblock} was then obtained by multiplying its count rate by the conversion factor of the block to which it belongs.
As shown in Figure \ref{fig:lum_dis}, during the $1.7\times10^6$ s observation by \ep-WXT from October 2024 to April 2025, the luminosity of $\sigma$ Gem varied between $7.2\times10^{30}$--$7.0\times10^{32}$ erg/s, spanning two orders of magnitude. The luminosity during identified flare blocks ranged from $3.0\times10^{31}$ erg/s to $7.0\times10^{32}$ erg/s, which was significantly higher than that during the pseudo-quiescent state. Meanwhile, the pseudo-quiescent state itself also exhibited a certain level of activity, with luminosity varying between $7.2\times10^{30}$--$5.2\times10^{31}$ erg/s (nearly one order of magnitude), and a time-weighted mean value of $2.1\pm0.6 \times 10^{31}$ erg/s. This scatter was significantly larger than the mean measurement error ($3.3 \times 10^{30}$ erg s$^{-1}$), indicating that persistent activity existed even during the pseudo-quiescent state when distinguishable flaring activity was absent, which also supports the statement in Section \ref{sec:flare_id}.

\begin{figure*}[h!]
\centering
\includegraphics[width=\hsize]{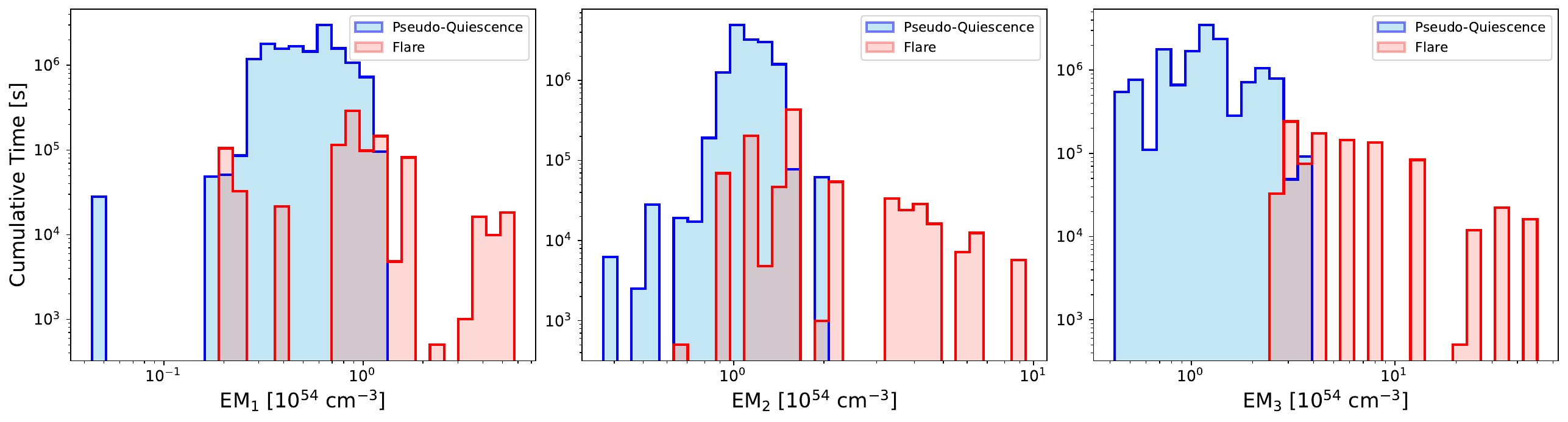}
\caption{Cumulative time spent at different EM levels for $\sigma$ Gem. The three panels (from left to right) show the distributions for EM$_1$, EM$_2$, and EM$_3$, respectively. The bar styles are the same as in Figure \ref{fig:lum_dis}.}
     \label{fig:em_dis}
\end{figure*}

Additionally, we performed a statistical analysis on EM values obtained from the fitting in Section \ref{sec:bw_ana} (Figure \ref{fig:em_dis}). For pseudo-quiescent state, the time-weighted mean values of the three EM components are $5.8\pm2.9 \times 10^{53}$ cm$^{-3}$ (mean error $1.1 \times 10^{53}$ cm$^{-3}$), $1.2\pm0.2 \times 10^{54}$ cm$^{-3}$ (mean error $1.2 \times 10^{53}$ cm$^{-3}$), and $1.7\pm1.2 \times 10^{54}$ cm$^{-3}$ (mean error $1.7 \times 10^{53}$ cm$^{-3}$), respectively. It is evident that among the three components, the hotter plasma exhibited larger EM values as well as greater variability. This trend reflects the sustained stellar activity during the pseudo-quiescence. 

\begin{figure*}[h!]
\centering
\includegraphics[width=16cm]{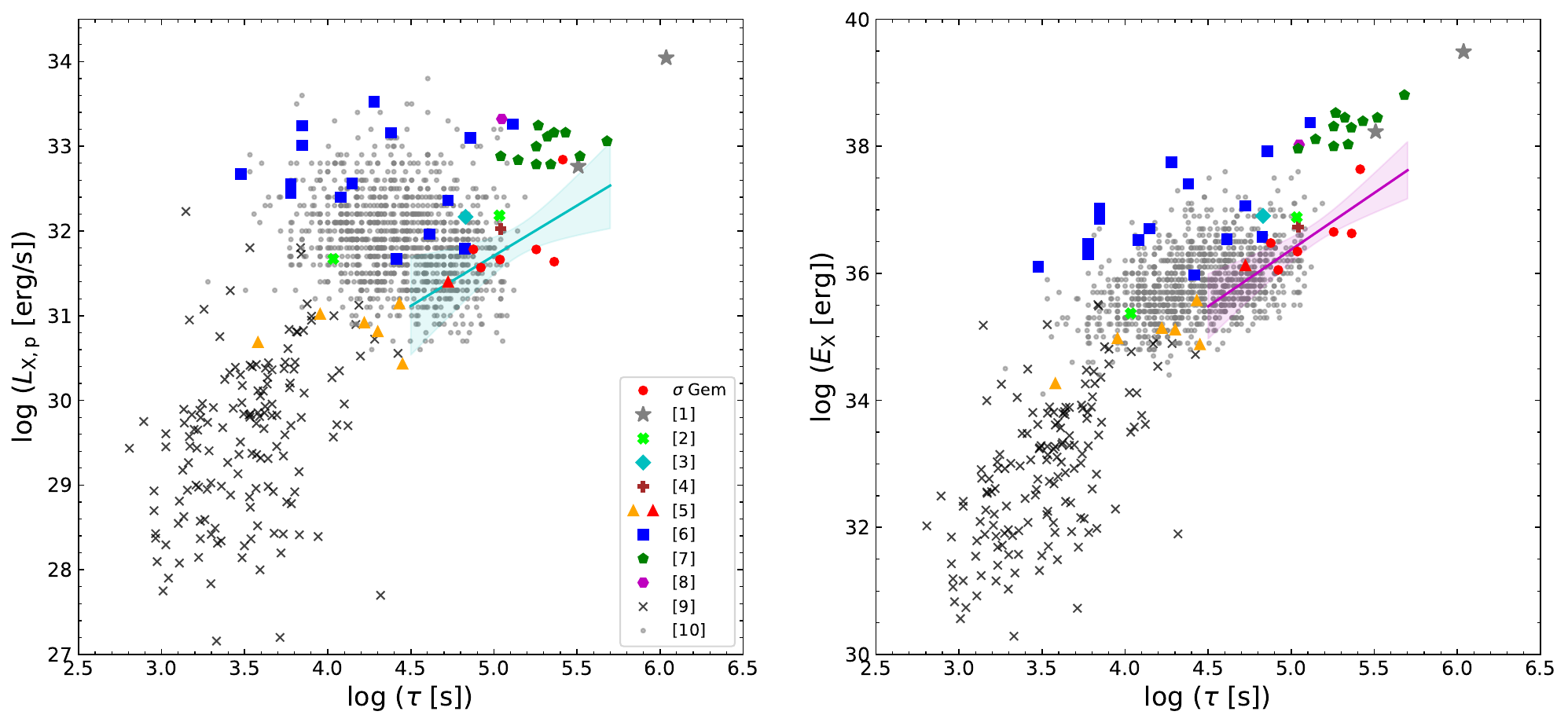}
\caption{Flare duration ($\tau$) vs. flare peak luminosity ($L_\mathrm{X,p}$, left panel) and flare energy ($E_\mathrm{X}$, right panel), along with stellar flares reported by [1]: \citet{Mao2025}. [2]: \citet{Tsuru1989}, [3]: \citet{Endl1997}, [4]: \citet{Franciosini2001}, [5]: \citet{Pandey2012}, [6]: \citet{Tsuboi2016}, [7]: \citet{Sasaki2021}, [8]: \citet{Karmakar2023}, [9]: \citet{Pye2015}, and [10]: \citet{Getman2021F}. A previous flare detected from $\sigma$ Gem is marked with a red triangle. The flares from [2] to [8] all took place in RS CVn-type binary systems, with the energy band converted into 0.5--4.0 keV. The energy bands are 0.2--12 keV in [9] and 0.5--8.0 keV in [10], respectively. The derived peak luminosity-duration and energy-duration relations for $\sigma$ Gem and their 1$\sigma$ uncertainty ranges are plotted in the left (cyan) and right (magenta) panels, respectively, as solid lines and shaded bands.}
     \label{fig:tau_lum_ene}
\end{figure*}

It should be noted that the EM values obtained here differ from those listed in Table \ref{tab:qui_fit}. The parameters in Table \ref{tab:qui_fit} were derived from fitting the combined spectrum of the pseudo-quiescent state and therefore represent an average condition, whereas the statistical analysis presented here more effectively captures the stellar activity.

\subsection{Flare characterization}
The derived properties of flares F1-F6 are listed in Table \ref{tab:energy_para}. The highest flux during each flare was adopted as the peak flux, denoted as $F_\mathrm{X,p}$. The fractional flux variation is defined as 
$f_{\mathrm{flare}} = (F_\mathrm{X,p}-F_\mathrm{X,q})/F_\mathrm{X,q}$, where $F_\mathrm{X,q}$ represents the pseudo-quiescent flux (see Section \ref{sec:qui}). For F1-F5, $f_{\mathrm{flare}}$ reached values around unity, indicating that the X‑ray emission produced locally during the flare peak was comparable to the total output of the entire star. In contrast, F6 showed a fractional variation of about 34.

In addition, the net flare energy released in the 0.5--4 keV band can be conveniently expressed as $E_\mathrm{X}=(L_\mathrm{X,p}-L_\mathrm{X,q})(\tau_\mathrm{r}/2+\tau_\mathrm{d})$, where the peak luminosity $L_\mathrm{X,p}=4\pi d^2 F_\mathrm{X,p}$. For F6, which showed a two‑stage decay, the energy was computed by $E_\mathrm{X}=(L_\mathrm{X,p}-L_\mathrm{X,q})(\tau_\mathrm{r}/2+\tau_\mathrm{d,1})+(L_\mathrm{X,T}-L_\mathrm{X,q})(\tau_\mathrm{d,2}-\tau_\mathrm{d,1})$, with $L_\mathrm{X,T}$ denoting the luminosity at the turning time $t_\mathrm{T}$.

To further investigate the relationship between X‑ray peak luminosity/energy release and flare duration, we compared our flares with those presented in Figure 4 of \citet[][ our Figure \ref{fig:tau_lum_ene}]{Mao2025}. This compilation includes flares from several RS CVn‑type systems \citep{Tsuru1989, Endl1997, Franciosini2001, Pandey2012, Tsuboi2016, Sasaki2021, Karmakar2023}, hundreds of flares on main‑sequence dwarfs \citep{Pye2015}, and the largest published sample of pre‑main‑sequence stellar flares to date \citep{Getman2021F}. Notably, the flares of $\sigma$ Gem occupy a parameter region consistent with other RS CVn systems, with both being substantially more luminous and longer‑lasting than flares on main‑sequence stars. While the peak luminosities of our flares are not exceptional among RS CVn flares, their durations are clearly longer than the 
average values of this class. This is likely driven by \ep's observational strategy, particularly its long time baseline, which is more capable of detecting long-duration flares.

For the six flares in this work, we performed a linear fit on their peak luminosity-duration and energy-duration relations in the logarithmic space, and the best-fit result is as follows:

\begin{equation}
\log_{10} (L_{\mathrm{X,p}}\mathrm{[erg\ s^{-1}]}) = (1.18\pm0.85) \times \log_{10} (\tau \mathrm{[s]}) + (25.8\pm4.4)
\end{equation}
\begin{equation}
\log_{10} (E_{\mathrm{X}}\mathrm{[erg]}) = (1.79\pm0.76) \times \log_{10} (\tau \mathrm{[s]}) + (27.4\pm3.9)
\end{equation}

The fitting results are plotted in Figure \ref{fig:tau_lum_ene} (cyan and magenta solid lines), with the shaded area representing the $ 1\sigma$ range. Notably, one of the six RS CVn-type stellar flares reported by \citet{Pandey2012} also originates from $\sigma$ Gem (the red triangle in Figure \ref{fig:tau_lum_ene}), whose spatial position is in close agreement with that derived in this work, thus confirming the consistency of the stellar activity of this object.

\begin{figure}[h!]
\centering
\includegraphics[width=\hsize]{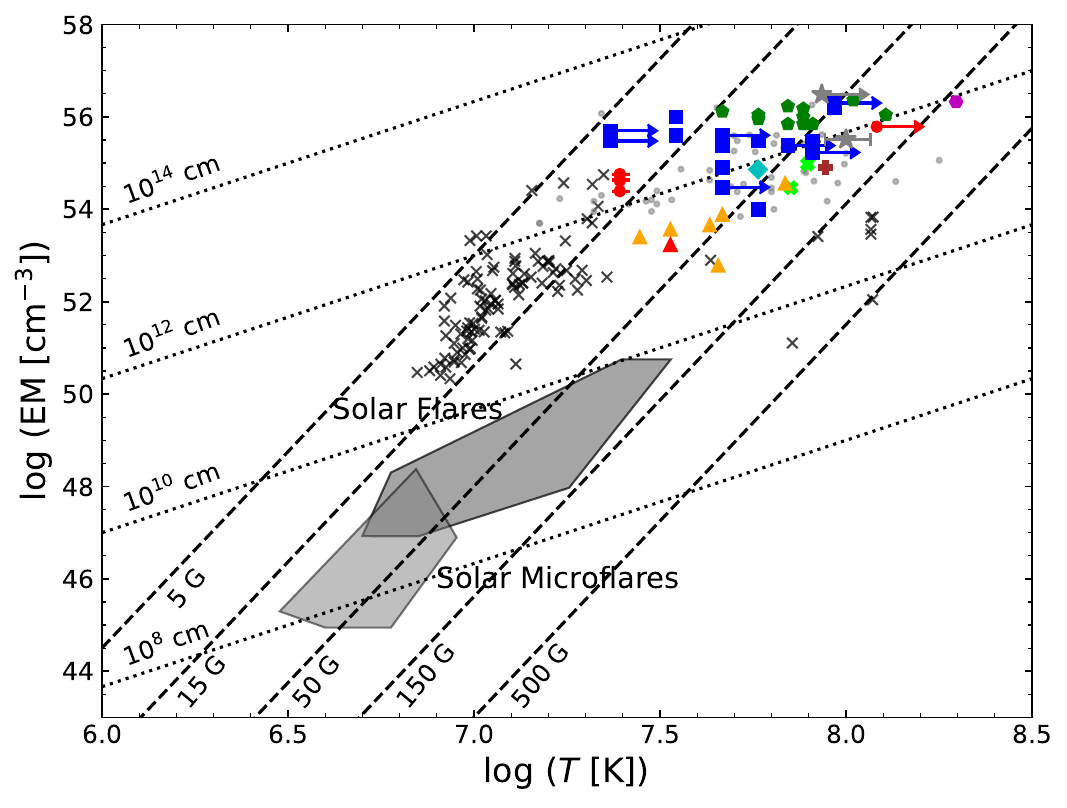}
\caption{EM vs. plasma temperature ($T$). Symbol styles follow Figure \ref{fig:tau_lum_ene}. Gray points, adopted from \citet{Getman2021}, represent a sample of 55 bright superflares originally identified by \citet{Getman2021F}. In addition, solar flares and microflares are denoted by filled polygons, using data from \citet{Feldman1995} and \citet{Shimizu1995}, respectively. Theoretical scaling relations are overplotted: the dashed lines correspond to the EM--$T$ dependence for a constant magnetic field (EM $\propto B^{-5} T^{17/2}$), while the dotted lines show the relation constrained by a fixed loop length (EM $\propto L^{5/3} T^{8/3}$), following \citet{Shibata1999}.}
     \label{fig:em_kt}
\end{figure}

Furthermore, we plotted these six flares on an emission measure-temperature (log EM-log $T$) diagram (Figure \ref{fig:em_kt}). For F1-F5, we adopted the temperature of the hottest plasma component, which was fixed at $kT_\mathrm{q,3}=2.12$ keV, and the corresponding peak emission measure (EM$_3$). For F6, the peak values of the fourth thermal component ($T_4$ and EM$_4$) were used instead. Following the scaling relations established by \citet{Shibata1999}, the characteristic magnetic loop lengths of the flares were estimated as $\sim10^{12}$ cm, corresponding to approximately 1.5 times the stellar radius. 
We note that these relations assume negligible heating during flare decay \citep{Reale2004}. While this assumption may hold for F1-F5, it is clearly violated for F6, whose decay light curve shows a distinct turning point. Therefore, the derived loop lengths should be regarded as upper limits. 

The magnetic field strengths were inferred to be around 10 G for flares F1-F5, while a lower limit of $\sim$100 G was derived for F6. Among flares reported on RS CVn stars (Figure \ref{fig:em_kt}), a $\sim$10 G magnetic field is relatively modest for such large flares, yet it is reasonable given the unremarkable flaring temperature and the sufficiently large flaring structure, following the relation $B \sim L^{-1/3} T^{7/6}$ derived from \citet{Shibata1999}. As reported by \citet{Tsuboi2016}, flares on RS CVn stars with even larger loop structures can have derived magnetic field strength as low as about 10 G, which is consistent with our results. The key to understanding why such magnetic fields can power strong flares lies in the flaring loop size: longer loops prolong the cooling timescale, resulting in longer flare durations and hence a larger integrated energy release.

\begin{figure}[h!]
\centering
\includegraphics[width=7cm]{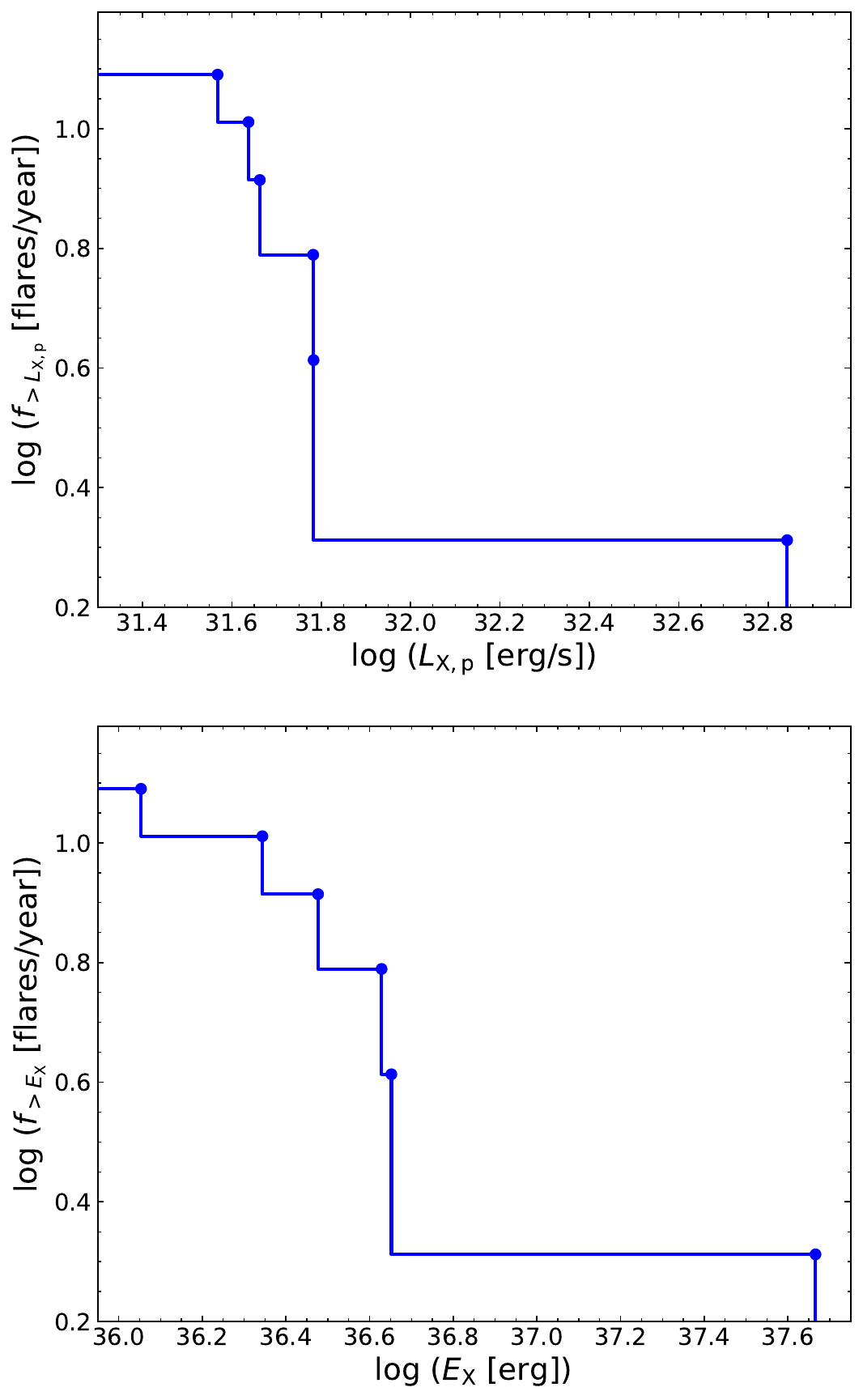}
\caption{The cumulative flare frequency distribution (FFD) of flare peak luminosity (top panel) and flare energy (bottom panel).}
     \label{fig:freq}
\end{figure}

\subsection{Flare frequency}

\ep-WXT data on $\sigma$ Gem were acquired between October 18, 2024, and April 13, 2025, during \ep's first year of operation. Observations were not feasible outside this interval due to Solar constraints. Within this $\sim$6-month window, a total valid exposure of approximately $1.69 \times 10^6$ s ($\sim$470 hours) was accumulated. Due to \ep's survey strategy and orbital constraints, these exposures were not continuous but were distributed across 986 ObsIDs, with no single GTI exceeding 3.35 ks.

This observational mode is a direct consequence of \ep's near-Earth orbit. The short orbital period is optimized for high-cadence sky monitoring, which significantly enhances the detection probability of X-ray transients. 
A corresponding limitation, however, is the short continuous on-target time (less than one hour per orbit) due to orbital effects such as Earth occultation and South Atlantic Anomaly (SAA) passages, which periodically interrupt the data stream during multi-orbit observations.

For the WXT data used in this work, the median observational gap was 4.2 ks, with a maximum of 356 ks. Only 0.8\% of the gaps exceeded 100 ks. In comparison, as listed in Table \ref{tab:energy_para}, the flares identified in the WXT data had durations ranging from 75 to 260 ks, with a mean value of 156 $\pm$ 72 ks. This mean duration was longer than most of the observational gaps. In other words, given this observational cadence, almost all long flares on $\sigma$ Gem, at least all flares with durations typical of F1-F6, should be temporally covered. To further validate the completeness of the flare detection, we performed a simulation (see Appendix \ref{sec:detect_simu}), confirming that the observational strategy can adequately detect flares comparable to F1–F6.

Therefore, we derived the flare frequency under the hypothesis that the detection of the six flares is complete within the half-year observational window. Figure \ref{fig:freq} presents the cumulative flare frequency as functions of both peak flare X-ray luminosity and flare energy.
As established in the literature \citep{Davenport2016, Olah2021, Getman2021F, Kriskovics2023, Odert2025}, the cumulative flare frequency distribution (FFD) is typically described by a broken power-law, characterized by two distinct slopes. The detected sample at the high-energy end is usually complete, exhibiting a steeper slope, while the slope becomes flatter toward the low-energy end due to the instrumental sensitivity limit preventing the detection of weaker flares. Our sample, however, is currently insufficient in size to reveal this characteristic double-slope feature. As more data accumulate over time, a sufficiently large sample will enable us to better constrain the FFD and thus provide a more reliable estimate of the flare rate at a given energy.

\subsection{Limitations}
Our analysis is subject to several instrumental and methodological limitations. First, while our calculation of net count rates accounts for the background-to-source region ratio and vignetting effects, it does not account for potential variations in detection efficiency among different CMOS detectors. Although all the 48 detectors were designed to exhibit identical responses, on-orbit performance degradation may introduce variations in detection efficiency across detectors and within individual pixels.

In this work, we have identified and excluded detections from known bad pixels and detector regions, with notable examples including affected areas in CMOS41 and CMOS34. However, unidentified bad pixels or regions may still affect the measured count rates. 
According to the full-scale calibration observation of WXT conducted in late 2025, the count rate discrepancies among different CMOS detectors for the same energy are within 8\%, with an average discrepancy of 4.0\% across the entire 0.5--4 keV band, indicating good uniformity. Nevertheless, we cannot fully rule out the presence of systematic artifacts in the light curves, which may introduce spurious change points in the Bayesian Blocks analysis.

\begin{figure}[h!]
\centering
\includegraphics[width=\hsize]{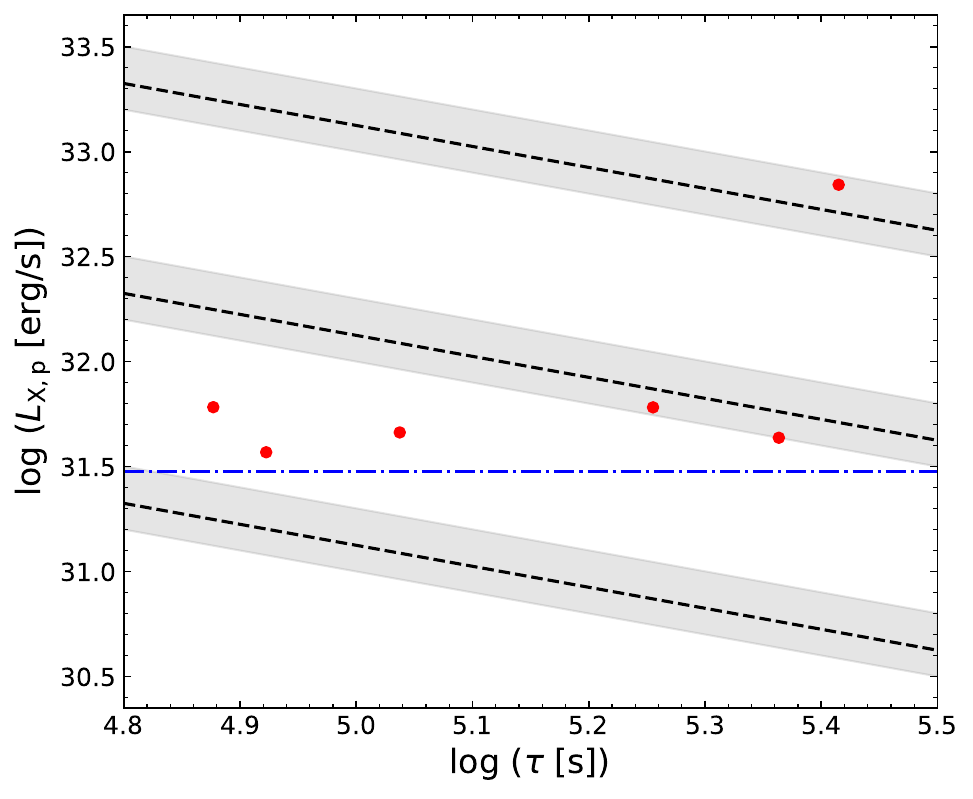}
\caption{Flare duration ($\tau$) vs. flare peak luminosity ($L_\mathrm{X,p}$) with iso-energy lines (dashed lines) with the assumption of equal rise and decay times. The shaded region bounds the iso-energy range, where the upper limit corresponds to a dominant rise phase($E_\mathrm{X}=\tau L_\mathrm{X,p}$), and the lower limit to a dominant decay phase ($E_\mathrm{X}=\frac{1}{2}\tau L_\mathrm{X,p}$). The blue dashdot line represents the detection threshold (minimum X-ray luminosity) for the flares.}
     \label{fig:tau_lum_lim}
\end{figure}

Furthermore, due to the sparse and uneven observational cadence, our data lack sensitivity to flux variations on short timescales. The flare identification method in our work is also inherently selective, as it can only identify flares with amplitudes exceeding $3\sigma$. For example, some visually flare-like structures in Figure \ref{fig:bblock}, such as the enhancement at the beginning of the light curve and the structure before flare F2, are excluded by this criterion. Their peak block count rates are above $R_\mathrm{q}+2\sigma(R_\mathrm{q})$, making them legitimate flare candidates, but they do not meet the $3\sigma$ threshold required for identification as flares. Another limitation is that, as shown in Figure \ref{fig:tau_lum_lim}, for flares with the same energy, shorter-duration flares are detectable only when their luminosities are sufficiently high. Finally, the method cannot distinguish flares from other stellar variability that may arise from distinct physical mechanisms.

\section{Conclusions}
\label{sec:conclu}
In this work, we presented an analysis of flaring activity on the active and X-ray bright RS CVn-type star $\sigma$ Gem, utilizing the first-year archival data from \ep\ mission. The WXT aboard \ep, with its exceptionally large FoV, performed high-cadence monitoring of the night sky. During the approximately six-month visibility window of $\sigma$ Gem, \ep-WXT accumulated a total exposure of $1.69\times10^6$ s. To select and extract WXT observations, we developed a data‑processing pipeline that produces a background-subtracted, vignetting-corrected light curve. Flares in the light curve were identified using the Bayesian Blocks method. Through simulations, we obtained the \texttt{\detokenize{ncp_prior}} to achieve a false positive rate of $p_0$ = 0.01 before applying the method to the actual observations.

The analysis revealed 6 significant ($\geq3\sigma$) flare events within this period. These flares were notably long-lasting, with durations between 21 hours and 3 days, exceeding those of most stellar X-ray flares. Their peak luminosities in the 0.5--4 keV band ranged from $3.7\times10^{31}$ to $7.0\times10^{32}$ erg s$^{-1}$, with total energies of $1.1\times10^{36}$ to $4.4\times10^{37}$ erg, placing them in the "superflare" category. Spectral analysis showed that the spectra could be well-fitted by a 3-T APEC model with the plasma temperatures fixed at the pseudo-quiescent values, with only the EMs allowed to vary. Meanwhile, the largest flare F6 exhibited distinct behavior: a fourth, hotter plasma component appeared around the flare peak, and its decay phase followed a double-exponential pattern. The estimated magnetic loop lengths required to produce these flares have upper limits of around 1.5 times the stellar radius.

Furthermore, parameters such as luminosity and plasma EM indicate significant ongoing activity even during non-flaring periods, justifying our use of the term "pseudo-quiescent" for these phases. We performed a statistical analysis of the flare frequency. Given that almost all the observational gaps for $\sigma$ Gem in the \ep-WXT data were shorter than the durations of the detected flares, we can reasonably conclude that our sample is complete for flares exceeding our $3\sigma$ detection threshold during this six-month window. Based on this complete yet small sample, we plotted the flare frequency distribution (FFD). With an extended observational baseline, a larger sample will allow us to better constrain the FFD and thus make a more reliable estimate of the flare occurrence rate at a given energy.

However, we acknowledge several limitations inherent to the \ep\ observational strategy and our methodology: \ep\ is optimized for long‑duration and high-amplitude flares, and we cannot distinguish flares unambiguously from other types of stellar activity.

$\sigma$ Gem was not known to have frequent flares before, as fewer than ten flares had been reported. Yet, within half a year of \ep\ observations, six flares were already detected, significantly expanding the flare sample for this star. With \ep\ continuously building up a vast archive of stellar X-ray data, we plan to apply our analytical framework to this growing dataset in the future. This will further enlarge the catalog of detected stellar X-ray flares, especially extreme events, and enable robust statistical studies of flare occurrence, energetics, and demographics across diverse stellar populations, thus substantially advancing our understanding of stellar flare physics.

\begin{acknowledgements}
X. M. acknowledges support from the University of Chinese Academy of Sciences (UCAS) for funding his participation in a training program at INAF - Osservatorio Astronomico di Palermo as a visiting scholar.
This work makes use of data obtained with Einstein Probe, a space mission supported by the Strategic Priority Program on Space Science of Chinese Academy of Sciences, in collaboration with the European Space Agency, the Max-Planck-Institute for extraterrestrial Physics (Germany), and the Centre National d'Études Spatiales (France). We acknowledge the support by National Key R\&D Program of China No. 2025YFF0511100, and the National Natural Science Foundation of China (Grant No. 12433005).

\end{acknowledgements}

\bibliography{sigGem}
\bibliographystyle{aa}

\begin{appendix}
\section{Observations by \ep-FXT}\label{sec:fxt}
The FXT consists of two co-aligned Wolter-I telescope modules (FXT-A and FXT-B), each equipped with a pn-CCD and identical in design to a single {\it eROSITA} telescope unit \citep{Predehl2021}. The system provides a FoV of about 1 square degree and operates in the 0.3--10 keV band, with an on‑axis spatial resolution of 20--24" (half‑power diameter, HPD). For a typical exposure of 10 ks, its sensitivity reaches $\sim1\times 10^{-14}$ erg cm$^{-2}$ s$^{-1}$.

\begin{table}[h!]
\caption{Observation log of \ep-FXT.}
\label{tab:fxt_log}
\centering
\setlength{\tabcolsep}{3pt}
\begin{tabular}{ccccc}
\hline\hline
No. & Setup & ObsID & Start Time & Exposure \\
 &  &  & (UTC) & ks \\
\hline
   01 & FF & 11908753667 & 2024-10-22 07:25:55 & 1.2 \\
   02 & FF & 11908778762 & 2024-10-29 16:52:09 & 1.1 \\
   03 & FF & 11908797700 & 2024-11-05 14:10:42 & 0.6 \\
   04 & FF & 11908821250 & 2024-11-12 11:26:28 & 1.1 \\
   05 & FF & 11908854536 & 2024-11-20 23:57:22 & 0.7 \\
   06 & FF & 11908888322 & 2024-11-27 00:13:45 & 0.7 \\
   07 & FF & 11908932864 & 2024-12-11 00:32:42 & 1.1 \\
   08 & FF & 11900016897 & 2024-12-24 09:11:26 & 1.2 \\
   09 & PW & 06800000484 & 2025-03-14 11:56:06 & 3.0 \\
\hline
\end{tabular}
\tablefoot{FF: Full Frame; PW: Partial Window.}
\end{table}

During the first year of \ep\ observation, \ep-FXT performed 9 ToO observations on $\sigma$ Gem (Table \ref{tab:fxt_log}). During the first eight observations, both FXT-A and FXT-B were conducted in Full Frame (FF) mode, while the ninth observation (ObsID: 06800000484) was performed in Partial Window (PW) mode. All nine observations were conducted with a thin filter. All X-ray data from both source and background regions were analyzed with the FXT data analysis pipeline (fxtsoftware v1.05). This processing produced cleaned event files, energy spectra, and corresponding response files. The \ep-FXT data are displayed along with \ep-WXT data in the top panel of Figure \ref{fig:fxt}. As shown in the figure, only the last FXT observation was performed during a flare epoch (F6), approximately two days after the F6 peak.

\begin{figure}[h!]
\centering
\includegraphics[width=\hsize]{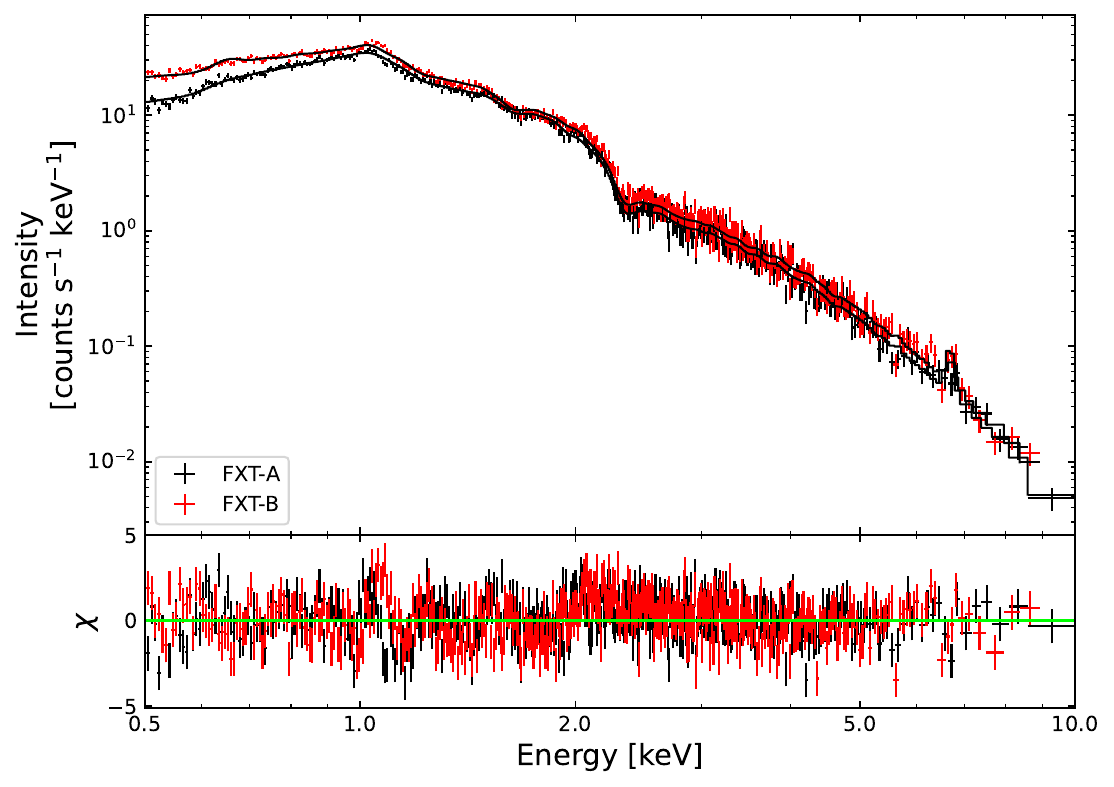}
  \caption{Joint spectral fitting for the 9th FXT observation. Black and red points represent the FXT‑A and FXT‑B data, respectively. The black solid line shows the best‑fit 3‑T APEC model.}
     \label{fig:fxt_spec}
\end{figure}

The 3-T APEC model was also used to perform joint spectral fitting of the FXT-A and FXT-B data from the same ObsIDs, with redshift fixed to zero and the metallicities of the three components tied together. All other parameters were left free. Compared to the earlier eight observations, the spectrum of the last one during F6 exhibited higher temperature in the cool components and increased EM in the hot components. This suggested that during this phase of the flare, the temperature of the flaring loop had declined to a level comparable to that of the coronal high‑temperature component. Furthermore, this spectrum showed no enhancement in either absorption or metallicity.

\begin{figure}[h!]
\centering
\includegraphics[width=\hsize]{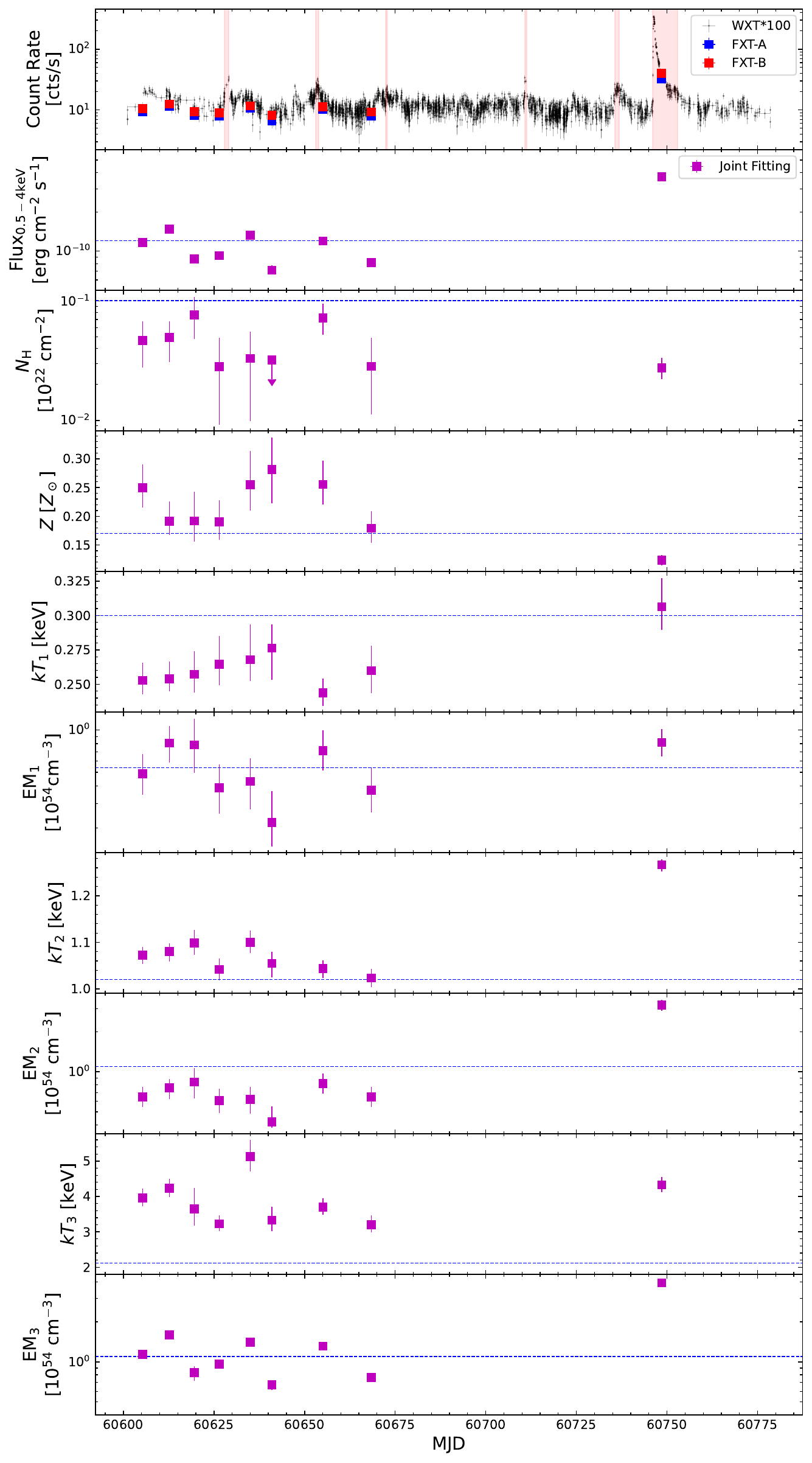}
  \caption{\ep\ light curve and FXT spectral parameters of $\sigma$ Gem. Panels from top to bottom display: 0.5--4 keV net count-rate, absorption-corrected flux, evolution of $N_\mathrm{H}$, $Z$, $kT_1$, EM$_1$, $kT_2$, EM$_2$, $kT_3$ and EM$_3$. In the top panel, black points represent WXT data (count rate scaled by 100 for display), while blue and red squares represent the FXT-A and FXT-B data, respectively. Flare intervals are marked in red (as in Figure \ref{fig:bblock}). In the remaining panels, the magenta squares denote the parameters from the joint fitting of FXT-A and FXT-B spectra, and the blue dashed lines correspond to the pseudo-quiescent values derived from WXT data fitted with 3-T APEC model.}
     \label{fig:fxt}
\end{figure}

We note that the spectral parameters obtained from the first eight FXT observations show a systematic offset relative to the pseudo‑quiescent values derived from WXT data (blue lines in Figure \ref{fig:fxt}). To rule out parameter degeneracy as the cause, we re-fitted the WXT spectra with the $N_\mathrm{H}$, $Z$ and temperatures fixed at their mean values from the first eight FXT observations. The poor quality of this fit, however, suggests that instrumental response differences, rather than degeneracy, may be the primary cause of the offset. Additionally, the extended high-energy bandpass of FXT compared to WXT provides a tighter constraint on the temperature of the hotter plasma component, $kT_3$.

\section{\texttt{\detokenize{ncp_prior}}-$p_0$ relation}\label{sec:ncp}
To investigate the relationship between \texttt{\detokenize{ncp_prior}} and $p_0$, we simulated 5000 source and 5000 background time series\footnote{In principle, more simulations yield a more stable and reliable relation, but at a higher computational cost. By adopting 5000 simulations, the threshold $p_0$=0.01 (50 out of 5000 trials) yields a relative uncertainty of $\sim$14\%, which we consider sufficiently robust for our analysis.}, respectively. In each simulated series, source (or background) events were uniformly distributed over the total summed Good Time Interval (GTI) duration, i.e., after removing the gaps between GTIs, with the total event counts matching the observed source and background counts. Each series was then binned with the same time bin width $\Delta t$ as used for the real data, which is 500 s, and the net count rate $R_{i,\mathrm{net}}$ was calculated for every bin following the procedure described in Section \ref{sec: pre_work}. Bayesian Blocks analysis was applied to each of the 5000 simulated light curves. Note that the start and end points of each time series are inherently detected as change points. Therefore, for a given \texttt{\detokenize{ncp_prior}}, we recorded the number of time series in which more than two change points were detected. This number corresponds to the count of false positive detections. The fraction of false positive detections among the 5000 simulations defines the false positive rate $p_0$.

We then varied \texttt{\detokenize{ncp_prior}} from 4 to 9 in steps of 1 and computed the corresponding $p_0$ for each setting. This yielded the relation shown in Figure \ref{fig:ncp_p1}. By linearly interpolating the curve, we obtained \texttt{\detokenize{ncp_prior}}($p_0$=0.01)=8.31.

\begin{figure}[h!]
\centering
\includegraphics[width=\hsize]{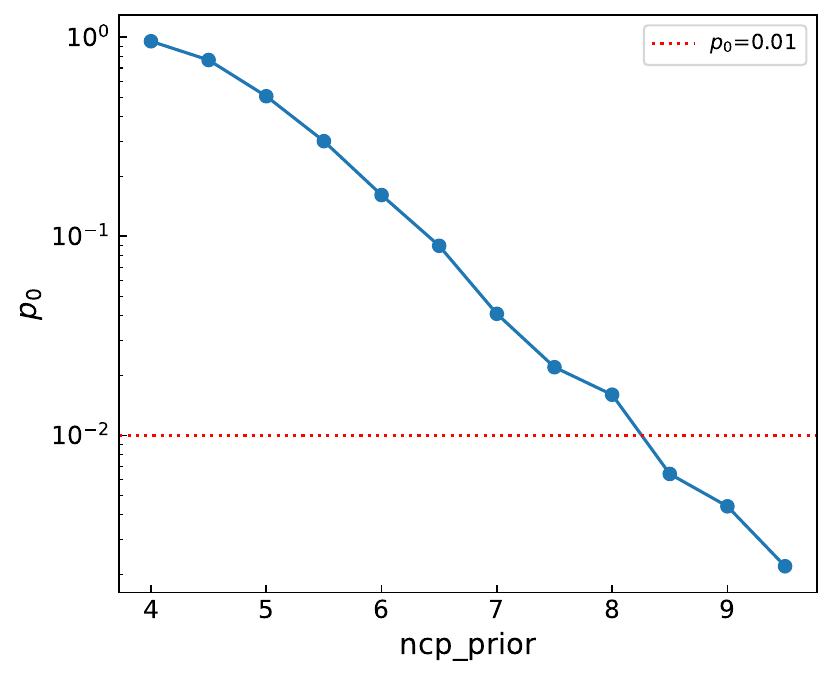}
\caption{False positive rate $p_0$ vs. \texttt{\detokenize{ncp_prior}} from simulation. The red dotted line marks the threshold of $p_0=0.01$.}
     \label{fig:ncp_p0}
\end{figure}

\section{Spectral parameters evolution of F1-F5}
\begin{figure*}[h!]
\centering
\includegraphics[width=18cm]{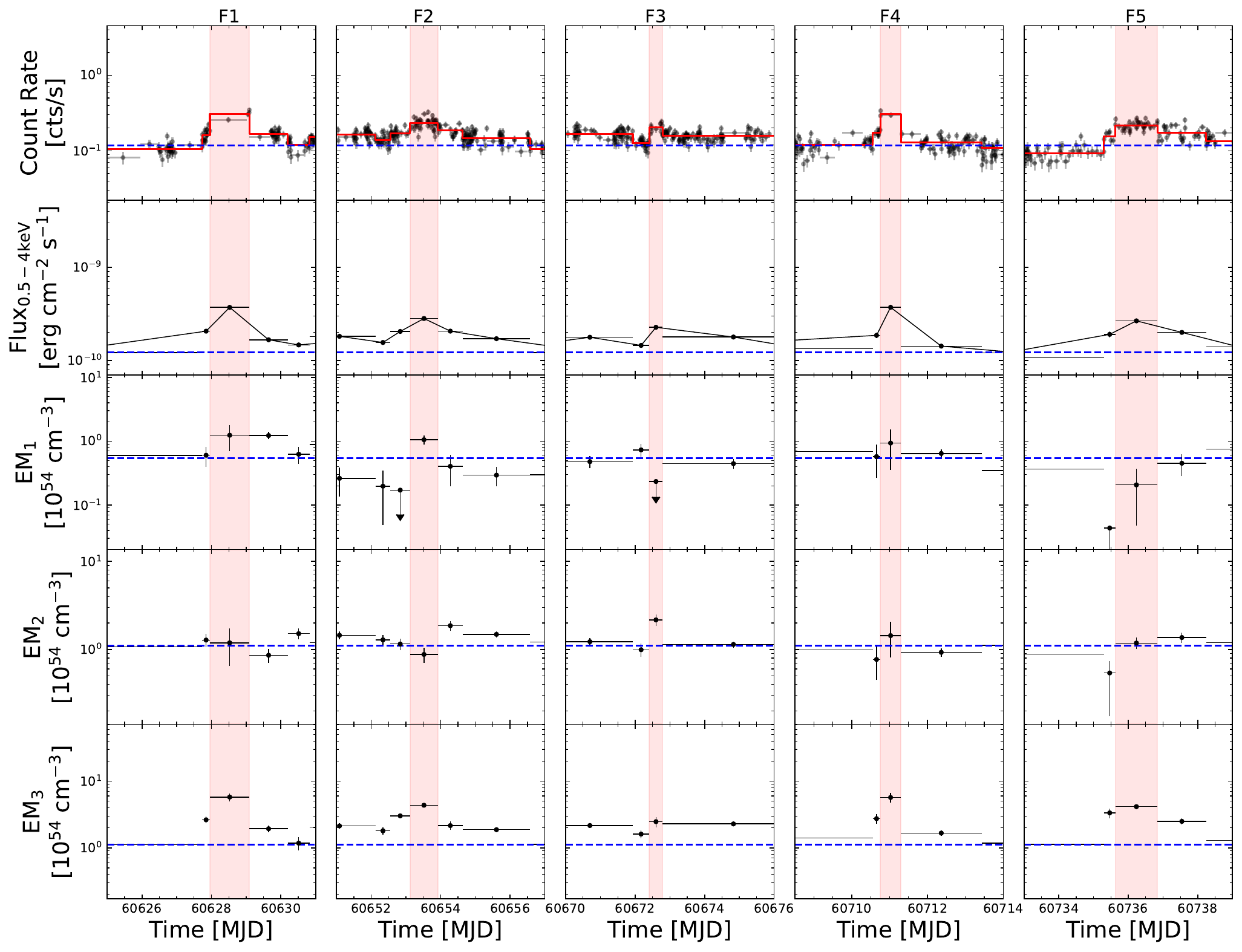}
\caption{X-ray light curves (top row) and temporal evolution of spectral parameters (lower rows) from 3-T APEC fitting for flares F1-F5, arranged from left to right. These panels are also the zoomed-in versions of the right panels shown in Figure \ref{fig:spec_evo}. The black points in the top panel represent the same vignetting-corrected net light curve shown in Figure \ref{fig:bblock}, determined in Section \ref{sec: pre_work}. The red shaded bands mark the flare blocks.}
     \label{fig:spec_evo2}
\end{figure*}

We present the temporal evolution of spectral parameters for flares F1-F5 in Figure \ref{fig:spec_evo2} as a complement to Figure \ref{fig:spec_evo}.

\section{Detection completeness}\label{sec:detect_simu}

\begin{figure*}[h!]
\centering
\includegraphics[width=8cm]{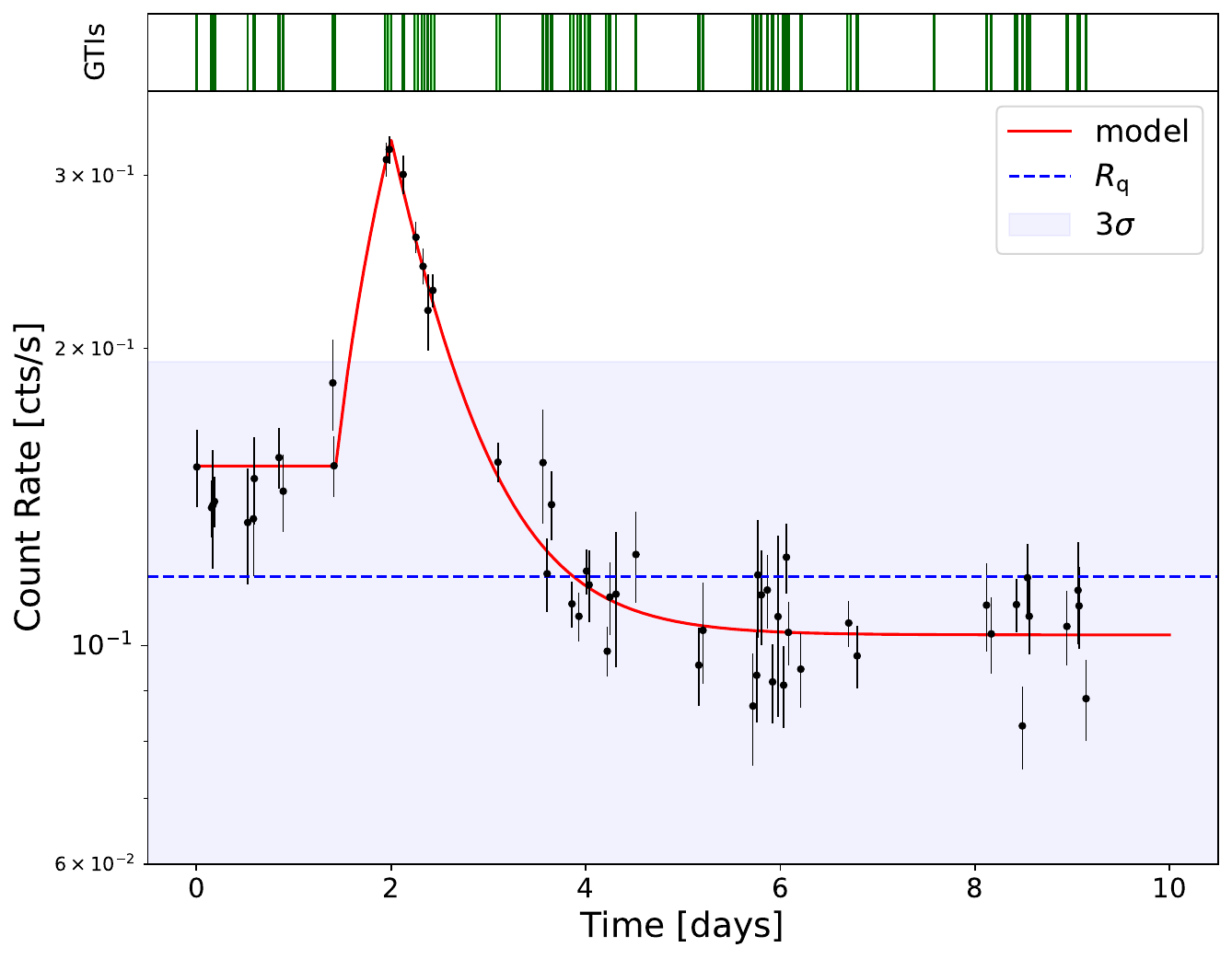}
\includegraphics[width=8cm]{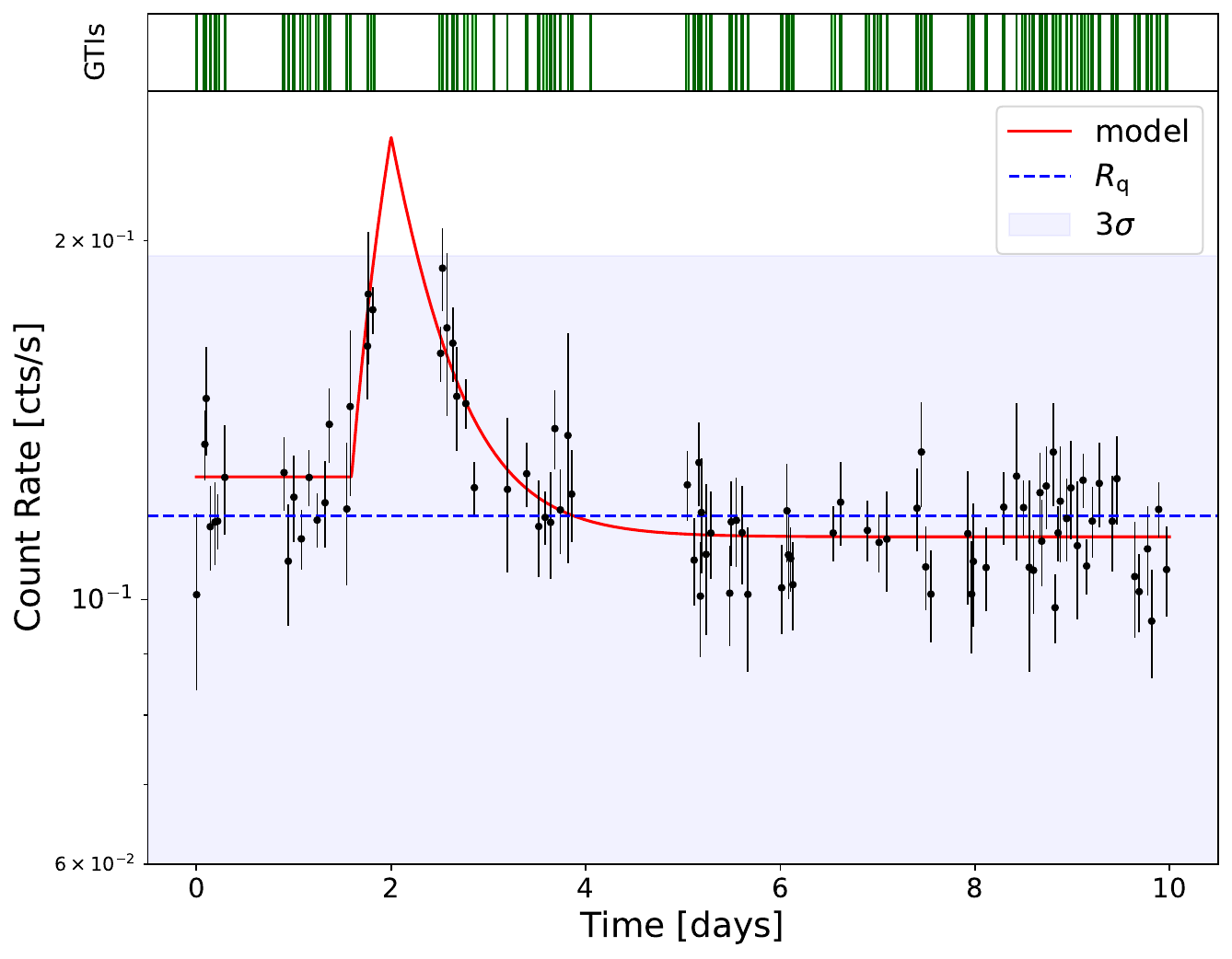}
\caption{Examples of simulated flare light curves. Left panel: a successful detection using the F2 template. Right panel: a failed detection using the F3 template. Black points are simulated data, one point per GTI.}
     \label{fig:detect_success_fail}
\end{figure*}

To investigate the adequacy of \ep's observational cadence for detecting a typical flare like F1-F6, we performed a simple simulation. We generated 1000 simulated light curves, each containing a single flare and spanning 10 days, long enough to fully capture the flare evolution. In each simulation, we randomly selected the fitted FRED template of one of the 6 flares in our work (red solid lines in Figure \ref{fig:lcurves}) and fixed the peak at $t_\mathrm{p}$ = 2 days. To mimic the observational cadence, we constructed a discontinuous time axis by alternating GTIs and gaps, sampling from their respective distributions shown in Figure \ref{fig:obs_stat} (d)-(e). For simplicity, we binned the light curve by GTI (each data point corresponds to a single GTI, as in Figure \ref{fig:detect_success_fail}) rather than by a fixed time width, with the counts per bin following a Poisson distribution. A flare was considered detected if any bin had a count rate exceeding $R_\mathrm{q}+3\sigma(R_\mathrm{q})$. Over 1000 simulations, the detection success rate was 93.5$\%$, high enough to support our statement that almost all long flares on $\sigma$ Gem should be detected by the observations. We therefore reasonably assumed that the 6 flares detected in our work are a complete sample during the 6-month window.

\end{appendix}

\end{document}